\documentclass{jcgt}

\usepackage{subcaption}
\usepackage{listings}
\usepackage{caption}
\usepackage{array, makecell}
\usepackage{changepage}
\usepackage{adjustbox}
\usepackage{pdflscape}

\setciteauthor{Majercik et al.}
\setcitetitle{Scaling Probe-Based Real-Time Dynamic Global Illumination for Production}

\accepted{2014-02-07}{2014-02-07}{2014-02-07}{Editor Name}{3}{1}{1}{1}{2014}
\seturl{http://jcgt.org/published/0003/01/01/}

\begin{document}
\title{Scaling Probe-Based Real-Time Dynamic \\ Global Illumination for Production}

\author
       {Zander Majercik
        \and Adam Marrs\\\\\hspace{18mm}NVIDIA
	\and Josef Spjut
       \and Morgan McGuire
       }

\teaser{
  \includegraphics[width=\columnwidth]{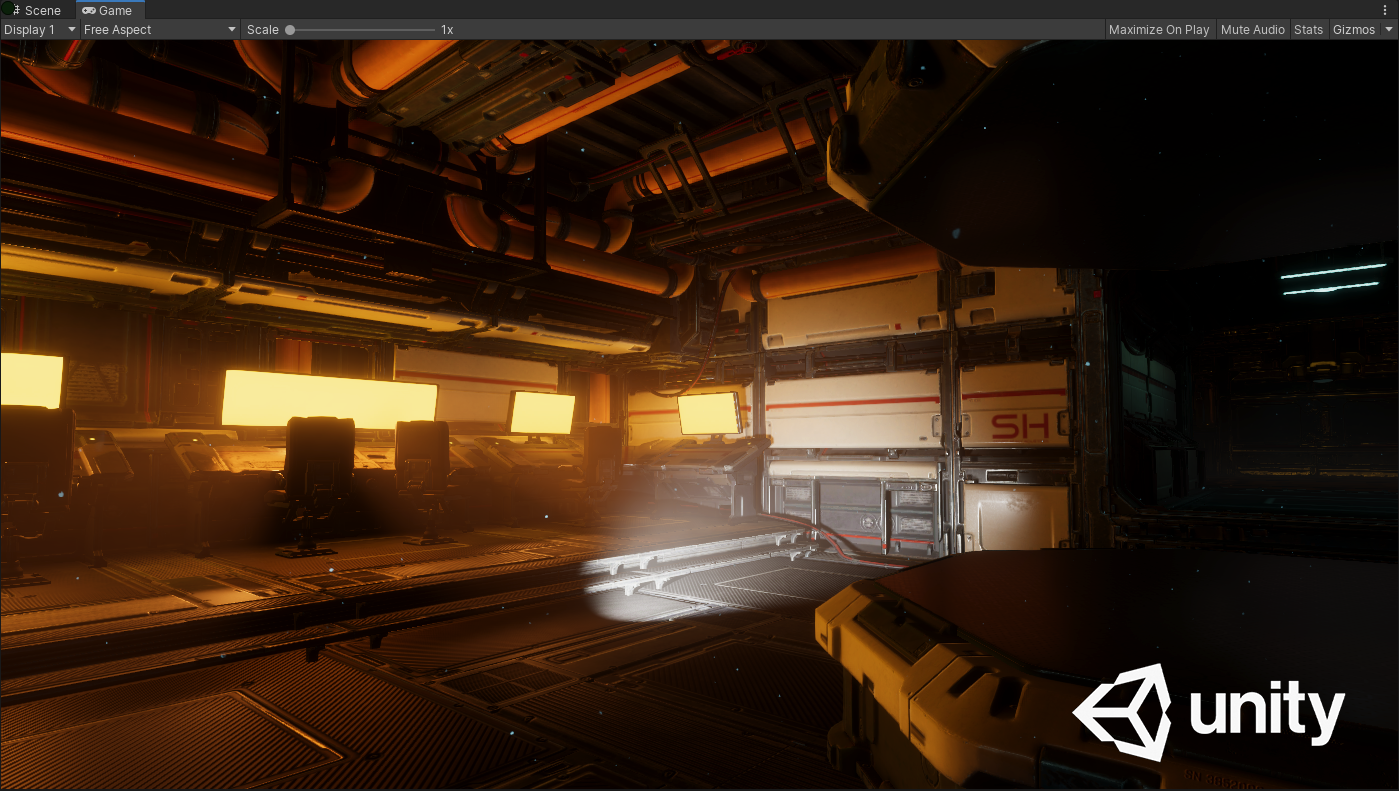}
  \label{fig:teaser}
  \caption{ Image rendered in a pre-release version of Unity with our global illumination technique. Most of the indirect lighting in this scene comes from emissives (the orange monitor screens) which are integrated automatically by our technique.}
}

\maketitle
\thispagestyle{firstpagestyle}

\begin{abstract}
\small
We contribute several practical extensions to the probe based irradiance-field-with-visibility representation~\cite{Majercik2019Irradiance}~\cite{Mara17Lightfield} to improve image quality, constant and asymptotic performance, memory efficiency, and artist control. We developed these extensions in the process of incorporating the previous work into the global illumination solutions of the NVIDIA RTXGI SDK~\cite{RTXGISDK}, the Unity and Unreal Engine 4 game engines, and proprietary engines for several commercial games. These extensions include: a single, intuitive tuning parameter (the ``self-shadow"  bias); heuristics to speed transitions in the global illumination; reuse of irradiance data as prefiltered \textit{radiance} for recursive glossy reflection;  a probe state machine to prune work that will not affect the final image; and multiresolution cascaded volumes for large worlds.
\end{abstract}

\section{Introduction}
\label{sec:intro}
This paper discusses an algorithm to accelerate the evaluation of global illumination.  The acceleration happens in two parts. The main part creates and maintains a data structure that allows a query of the form \emph{irradiance(location, orientation)} ($E(X, \omega)$), which replaces a potentially expensive computation of diffuse global illumination with a O(1) lookup into a data structure for locations anywhere in space. The second part re-uses that data structure to sample weighted average of incident radiance for glossy global illumination ($\int_\Gamma L(X, \omega) \cdot W(X, \omega) d\omega$) and combines the result with filtered screen-space and geometric glossy ray tracing.

This paper describes a refinement of a previous version of the diffuse portion of this method~\cite{Majercik2019Irradiance}.  This refinement is the union of what we learned when incorporating that algorithm into several products, including the Unity game engine, the Unreal Engine 4 game engine, the NVIDIA RTXGI SDK version 1.1~\cite{RTXGISDK}, and several unannounced commercial games. These learnings include changes to the underlying algorithm to improve quality and performance, advice on tuning the algorithm and content, expansion of the algorithm to a complete solution that also accelerates glossy reflections, and system integration best practices for these methods. This was driven by constraints from various platforms, requests from game developers and game artists, and new research on the problem. Because they were developed across several different productization efforts with different vendors, we believe that these learnings are fairly universal and robust, but they should not be construed as describing the features or performance of any one in particular.

A key element of our algorithm is a probe, which stores directional information at a point.  Environment maps are a type of probe---they store distant radiance as seen from any point in the scene.  Our probes store irradiance, weighted averages of distance, and weighted averages of squared distance (See Table~\ref{tab:definitions} for terms we use in relation to probes) for a 3D grid-like structure of points in the scene.

\begin{table}[htb]
\begin{adjustwidth}{-2cm}{}
  \begin{tabular}{p{4cm} p{12cm}}
 \textbf{Term} & \textbf{Definition} \\
\hline
Probe & A probe stores data at a point with values for directions on the sphere.  \\
Probe Query & Trilinear interpolation (bilinear filtering and direction) and a visibility and angle weighted interpolation between multiple probes. The net result is an irradiance value that esimates the irradiance field at a point relative to a normal. \\
Irradiance & Incident power per unit area; the cosine-weighted integral of radiance relative to the sample direction.  \\
Weighted sum of distance & Weighted sum (a weighted average in our implementation) of the distance to the nearest surface seen from a 3D point in a particular direction. In our case we use a cosine raised to a power. \\
Direct lighting & Light that is emitted from a light source, reflects from one surface, and then reaches the viewer.  \\
Indirect lighting &  Light that reflects off two or more surfaces before reaching the viewer (all lighting that is not direct) \\
Global illlumination & Light that includes both direct and indirect lighting. \\
  \end{tabular}
  \caption{ Terms and definitions. }
      \label{tab:definitions}
\end{adjustwidth}
\end{table}

Our algorithm has several components related to the organization, computation, and querying of probes.  The new information described in this paper is indicated in Table~\ref{tab:algtable}.  In the table, we indicate what is new relative to descriptions of previous versions of this algorithm. In addition, we give a complete description of the full algorithm below so that readers will not need to consult descriptions of previous versions to understand the algorithm.

\section{Overview of the algorithm}

At the core of the algorithm are probes that store weighted sums of color, distance, and squared distance.  A 2D version of a probe storing a weighted average of distance to nearest object is shown in Figure~\ref{fig:rayblending}.  These probes are processed as follows.

\subsection{Build and Initialization}
Start by building a 3D grid. From that grid, optimize probe positions by moving them outside of static geometry (Section~\ref{sec:OP}). Then, classify all probes into ``Off'',``Sleeping'', ``Newly Awake'', ``Newly Vigilant'', ``Awake'', or ``Vigilant'' (Section~\ref{sec:probestates}). At the end of this stage, all probes are in their final positions and initial states.

\subsection{Probe Query}
Take a 3D point (within the probe volume) and normal direction. For every point within the volume, there are 8 probes (corners of a 3D box) that surround it. Loop over those 8 probes. For each one, compute a weight based combination of:
\begin{itemize}
\item trilinear weight from probe position
\item backface weight (is the probe behind the point relative to the normal?)
\item visibility (can the probe see the point?). This includes a self-shadow bias' term for robust occlusion queries (Sec.~\ref{sec:selfshadowbias}).
\end{itemize}
Sample the value from each probe in the direction of the normal, and sum those using the computed weights. That is the sampled irradiance value.

For multiple volumes, do this for each volume, and then weight between the volumes as described in Section~\ref{sec:multipleprobevolumes}. Volume blending with tracking windows is discussed in Section~\ref{sec:cameratrackingvolume}.

\subsection{Probe Update}
\begin{figure}[htb]
  \centering
   \includegraphics[width=0.9\columnwidth]{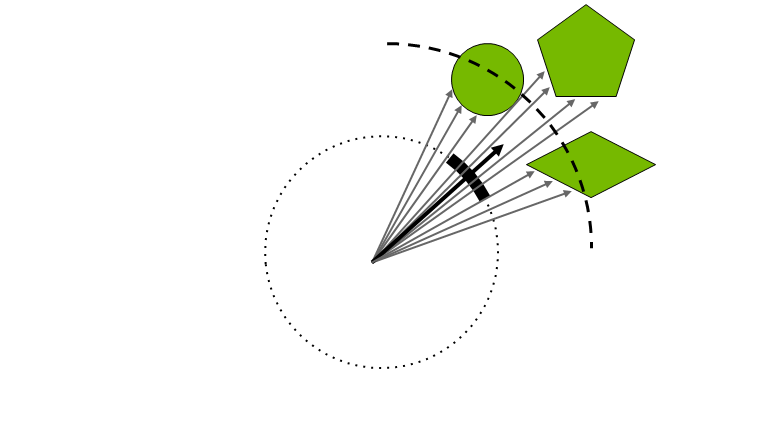}
   \caption{\label{fig:rayblending}
A 2D probe for illustration. This probe shows one ``cell'' (texel) as the bold segment of the circle. The bold arrow is the direction associated with the cell. The cell stores the weighted average of the hit distances of each of the sample directions. Note that this weighted average includes directions ``outside'' the center cell. The weighting function is larger for directions near the cell center, and the resulting weighted average is thus influenced more by the longer directions in this particular example. The bold dotted line is the stored ``distance'' in the cell. Note that a direction can contribute to more than one cell, and we loop over directions updating any cell that a direction contributes to.}
\end{figure}

For each probe that is ``Awake'' or ``Vigilant'' (Section~\ref{sec:probestates}), trace rays in a spherical fibonacci pattern, rotating the pattern randomly every frame. Shade these ray hits using the normal deferred shading algorithm, including sampling the probe volume to include the irradiance from the probes. A section of an example ray cast, with a texel to which the rays contribute highlighted, is shown in Figure~\ref{fig:rayblending}. The update then proceeds for both irradiance and mean distance values as follows.

\paragraph{Irradiance}
Compute a cosine-weighted average of the radiance values of these shaded ray hits relative to the direction of each probe texel. Then, for each probe texel, blend these newly computed values into the probe texel at a rate of $(1 - \alpha)$---we refer to this alpha term as ``hysteresis''. We adjust this hysteresis per probe and per texel based on our convergence heuristics, described in Section~\ref{sec:heuristics}.

\paragraph{Mean Distance and Mean Distance-Squared}
Compute a \emph{power}-cosine weighted average of the distance values for each ray relative to the direction of each probe texel. For each probe texel, blend these values as with irradiance above. We adjust the hysteresis for mean distance separately from irradiance---details and reasoning are provided in Section~\ref{sec:heuristics}.

We update the probe texels by alpha blending in the new shading results at a rate of $1-\alpha$, where $\alpha$ is a \textit{hysteresis} parameter that controls the rate at which new irradiance and visibility values override results from previous frames (Eq. \ref{eq:1}). We dynamically adapt this hysteresis value per-probe and per-texel (Section~\ref{sec:heuristics}). The upodate equation is as follows:

\begin{equation}
\begin{aligned}
&E'[\hat{n}] =  \alpha E[\hat{n}] + (1-\alpha) \!\!\!\!\sum_{\textrm{ProbeRays}}\max(0, \hat{n} \cdot \hat{\omega}) \cdot L(\hat{\omega})
\end{aligned}
\end{equation}

Where $E$ is the old irradiance/visibility texel in direction $\hat{n}$, $E'$ is the new texel value, $\hat{\omega}$ is the direction of the ray, and $L(\hat\omega)$ is the radiance transported along the ray.

\section{Related Work}
\label{sec:relatedwork}

Interactive global illumination has been an active area of research for years. We review the areas most relevant to our work.

\paragraph*{Interactive Global Illumination with Light Probes}

Image-based lighting solutions are ubiquitous in modern video games~\cite{Martin2010Radiosity,Ritschel:2009:ADG:1507149.1507161,FarCry2012,Hooker2016GI}. A common workflow for such solutions involves placing light probes densely inside the volume of a scene, each of which encodes some form of a spherical (ir)radiance map. Prefiltered versions of these maps can also be stored to accelerate diffuse and glossy runtime shading queries.

Variants of traditional light probes allow artists to manually place box or sphere proxies in a scene. These proxies are used to warp probe queries at runtime in a manner that better approximates spatially-localized reflection variations~\cite{Sebastien2012Parallax}. Similarly, manually-placed convex proxy geometry sets are also used to bound blending weights when querying and interpolating between many light probes at runtime, in order to reduce the light leaking artifacts common to probe-based methods.

Practitioners agree that eliminating manual probe and proxy placement remains an important open problem in production~\cite{Hooker2016GI}. Without manual adjustment of traditional probes, it is impossible to automatically avoid probe placements that lead to light and dark (i.e., shadow) leaks or displaced reflection artifacts. Majercik et al.'s~\shortcite{Majercik2019Irradiance} light probes avoid light and dark leaking with raytraced visibility information, but placing these probes in a uniform grid still leads to suboptimal probe locations (e.g. probes stuck in walls). To avoid these issues for glossy GI, some engines rely instead on screen-space ray tracing~\cite{Valient2014SSRR} for pixel-accurate reflections. These methods, however, fail when a reflected object is not visible from the camera's point of view, leading to inconsistent lighting and view dependent (and so temporally unstable) reflection effects. 

\begin{landscape}
\begin{table}[h]
\centering
\scriptsize
  \begin{tabular}{| p{2.5cm} | p{4cm} | p{4cm} | p{4cm} | p{4cm} | }
\hline
  & 
Previous Approaches 
\begin{itemize}\setlength{\itemsep}{-5pt}\setlength{\itemindent}{-30pt}
\item[] \cite{Martin2010Radiosity}
\item[] \cite{Ritschel:2009:ADG:1507149.1507161}
\item[] \cite{FarCry2012}
\item[] \cite{Hooker2016GI} 
\item[] \cite{GIDivision} 
\end{itemize}
& Light-field probes 
\begin{itemize}\setlength{\itemsep}{-5pt}\setlength{\itemindent}{-30pt}
\item[] \cite{Mara17Lightfield}
\item[] \cite{FastNonUniformPlacement19} 
\end{itemize}
& DDGI
\begin{itemize}\setlength{\itemsep}{-5pt}\setlength{\itemindent}{-30pt}
\item[] \cite{Majercik2019Irradiance} 
\end{itemize}& This work \\
\hline
Spatial Organization & 3D grid, with manually placed probes and box proxies, algorithmically precomputed probe locations & 3D grid, $4\times4\times4$ probes~\shortcite{Mara17Lightfield} Non-uniform automatic placement over static geometry~\shortcite{FastNonUniformPlacement19} & 3D grid,varying resolutions & 3D grid with offsets, multiple volumes, tracking windows \\
\hline
Encoding  & Cube maps & Octahedral encoding~\cite{Cigolle2014Vector}, $1024\times1024$ & Octahedral, varying resolutions & Octahedral, $8\times8$ irradiance, $16\times16$ visibility \\
\hline
Initialization & Precomputed & Static, precomputed & Uniform initialization to 0, value converges with update &Classified into states based on update rate, converge “live” probes \\
\hline
Update & Static. Dynamic lighting, static geo~\cite{GIDivision} & Static, precomputed &  Ray trace with alpha blending, pixel shader with stencil buffer & Ray trace with dynamic alpha blending (convergence and perception), Optimized convolution compute shader \\
\hline
Query & Shading weights based on manually placed proxy geometry & Light field ray tracing using probes & Raster for direct lighting. Use WS positions to query 8 probe cage with variance bias, chebyshev bias, loads of bias terms & Previous probe sampling +  single bias term (self-shadow bias), multivolume blending, primary hit glossy raycast + second order glossy reflection sampled from probes \\
\hline
  \end{tabular}
  \caption{ Evolution of probe based GI showing spatial organization, encoding, initialization, update, and query for the GI computation.}
      \label{tab:algtable}
\end{table}
\end{landscape}
\clearpage

\textit{Light Field probes}~\cite{Mara17Lightfield} automatically resolve many light/dark leaking issues (in scenes with static geometry and lighting) by encoding additional information about the scene geometry into spherical probes. A solution for dynamic lighting is presented in Silvennoinen et al.~\shortcite{Silvennoinen:2017}, but this solution only supports coarse dynamic occluders and requires complex probe placement based on static geometry.  As mentioned above, the irradiance probes of Majercik et al.~\shortcite{Majercik2019Irradiance} avoid most light/dark leaks in scenes with dynamic lighting and geometry, but probe placement is stilll suboptimal. Suboptimal placement can lead to lighting results that, while believable, are inferior to the correctly sampled result, and sometimes exhibit shadow leaking in cases of complex geometry with acute corners.

\paragraph*{Interactive Ray Tracing and Shading.}

Correct shading with probe-based lighting methods relies on point-to-point visibility queries. At a high-level, one can interpret our ray tracing technique as tracing rays against a voxelized representation of the scene (as in voxel cone tracing), but with a spherical voxelization instead of an octree. Two important differences that contribute to many of the practical advantages of our representation are 1) we \textit{explicitly} encode geometric scene information (i.e. radial depth and depth squared) instead of relying on the \textit{implicit} octree structure to resolve local and global visibility details, and 2) that neither our spatial parameterization nor our filtering relies on scene geometry. This prevents light (and dark) leaking artifacts and allows us to resolve \textit{centimeter-scale} geometry at about the same cost (in space and time) as a voxel cone tracer that operates at \textit{meter-scale}. As we target \textit{true world-space} ray-tracing in a pixel shader, and not just screen-space ray tracing, our technique can be seen as a generalization of many previous, e.g., real-time environment map Monte Carlo integration methods\cite{Stachowiak2015Stochastic,Wyman2005Refraction,Toth2015Disparity,Jendersie_jcgt16_RTGI} .

\paragraph*{Probe Representation.}

As in the work by Majercik et. al~\shortcite{Majercik2019Irradiance}, we apply Cigolle et al.'s~\shortcite{Cigolle2014Vector} octahedral mapping from the sphere to the unit square to store and query our spherical distributions. This parameterization has slightly less distortion than cube maps and provides easier methods for managing seams. In this work, we select resolutions for octahedral irradiance and mean distance/distance squared for quality and performance.

\paragraph*{GI in Production: A Motivating Example}

In both offline and realtime rendering, significant previous work has been devoted to adapting existing global illumination algorithms for production. Path tracing in film, which radically changed both artist workflow and render farm computation load, is a good example. The core path tracing algorithm has remained largely unchanged, but practical considerations of the particular hardware and software systems required specialized updates to the technique~\cite{keller2015path}.

Similarly, our extensions to the previously published DDGI algorithm are a guide for adapting it and other probe-based techniques to a production setting. We report real changes that we made to the base algorithm to fit production constraints.

\section{Qualitative Image Improvements}
\label{sec:contributions}

\subsection{Self-shadow bias for correct visibility}
\label{sec:selfshadowbias}

\begin{figure}[htb]
  \centering
   \includegraphics[width=0.5\columnwidth]{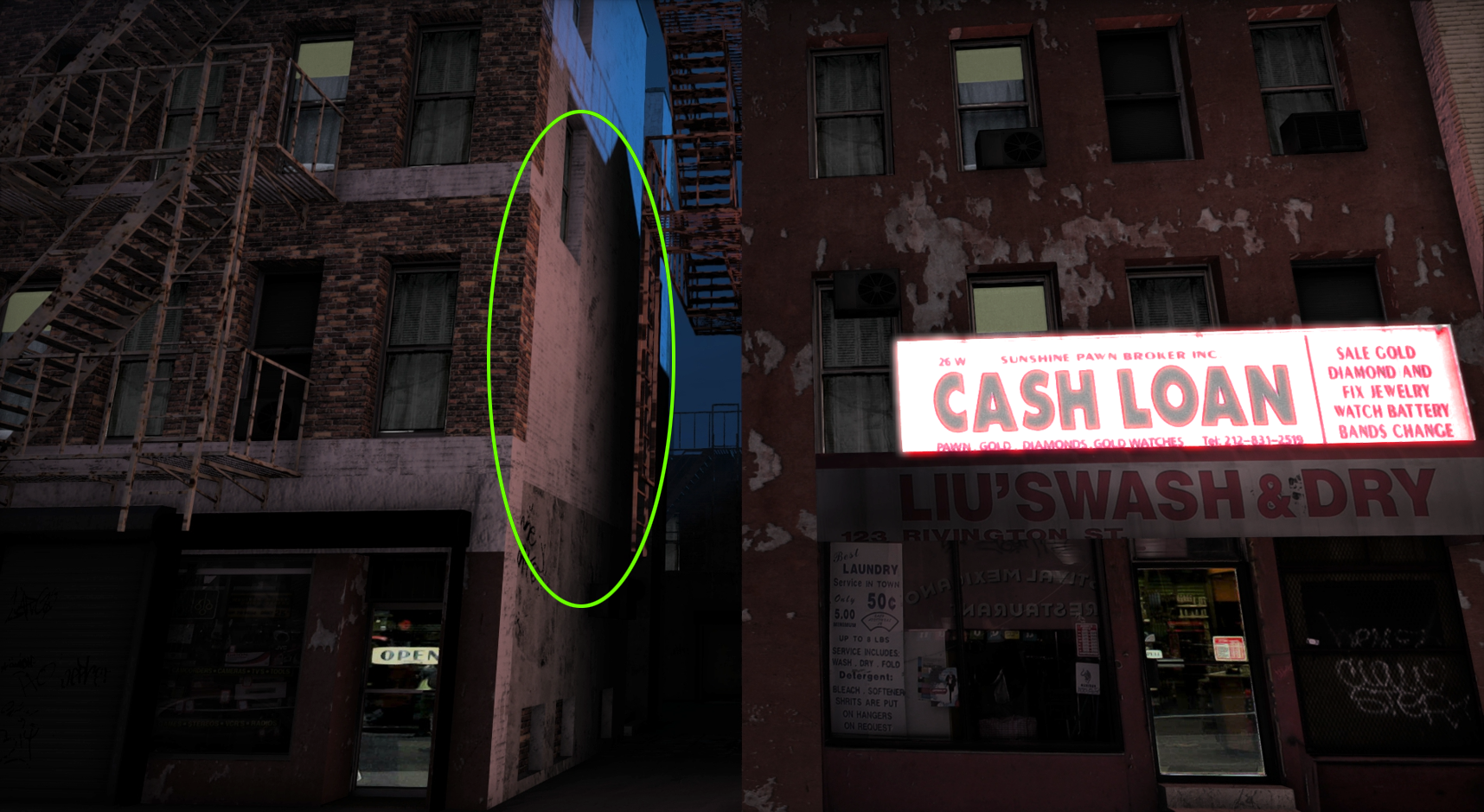}~\includegraphics[width=0.5\columnwidth]{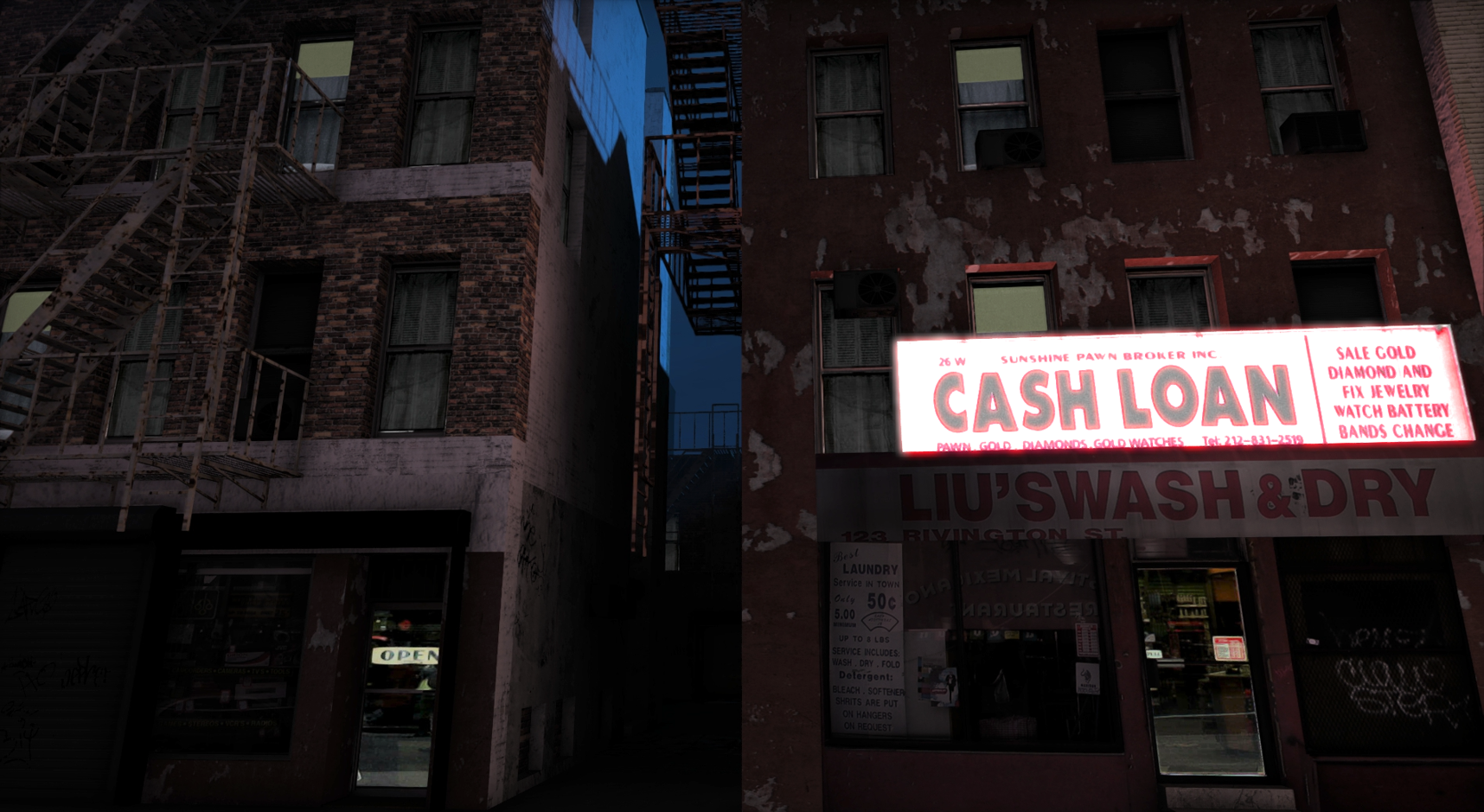}
   \caption{\label{fig:selfshadowbias}
A night scene from our prototype. The wall entering the alley in the left image shows light leaking due to overly high self-shadow bias. The correct self-shadow bias in the right image computes proper occlusion.
}
\end{figure}

When querying the probe volume at a surface, variance in the visibility estimate will be highest around the mean of the distribution---in other words, \emph{at} the surface (see Figure~\ref{fig:selfshadowbias}). To avoid the shadow leaking that results from this, an additional bias away from the mean of the distribution is added to the sample point during probe query. The previous technique~\shortcite{Majercik2019Irradiance} used a combination of scene-tuned biases on the mean of the distribution, the variance of the distribution, and the chebyshev statistical test to move the visibility query to a point of lower variance in the distribution. 
Intuitively, "a point of lower variance in the distribution" can be thought of as a point slightly offset from the surface (in world space). Thus, we unify these statistical bias parameters into a single \textit{self-shadow} bias term. The self-shadow bias is a world-space vector pointing away from the initial sample point on the surface and is computed as follows:

\begin{equation} \label{eq:1}
\begin{aligned}
&BiasVector =  (\mathbf{n} * 0.2 + \omega_o * 0.8) * ( 0.75 * D) * B
\end{aligned}
\end{equation}

Where $\mathbf{n}$ is the normal vector at the sample point, $\omega_o$ is the direction from the sample point to the camera, $0.2$ and $0.8$ are empirically determined constants, $D$ is the minimum axial distance between probes, and $B$ is a user-tunable floating point scalar. We add this bias vector to the initial sample point to yield a new point which we use for the visibility test.

Our self-shadow bias is more robust than the previous biases because a default value of the $B$ parameter (0.3f) worked well for most scenes, whereas the previous biases each had to be specifically tuned per scene. In cases where scene specific tuning is necessary, tuning is easier because we present a single tunable parameter instead of three. Generally, a higher self shadow bias is necessary when there is increased variance in the depth estimate, as would be the case when lower ray counts are used to update the probes (as might be done to improve performance).

To further decrease light leaking, probe update rays that hit backfaces record a value of 0 for irradiance and shorten their depth values by 80\%. Shortening depth values ensures that the probe will see backface surfaces as shadowed and not light them. We set irradiance to 0 to ensure that any lighting that does come from that probe does not cause light to leak where it should not. We do not set depth values to 0 for two reasons: 1) it would drive the computed chebyshev weight towards 0, which might be driven higher when the weights are normalized and 2) probes that see some backfaces but are not stuck in walls (due to idiosyncrasies of geometry) could have overly skewed average depths if many of them were set to 0.

To minimize the number of probes stuck in walls as much as possible, we offset probe positions using an iterative adjustment algorithm, as described in Section~\ref{sec:OP}.

\subsection{Perception-based exponential encoding}
 If the irradiance probes are slow to converge, abrupt lighting changes in a scene can create noticeable lag in the diffuse indirect illumination. The lag is most salient in light-to-dark transitions.
To combat this, we accelerate convergence by applying a perception-based exponential gamma encoding to probe irradiance values. This encoding interpolates \emph{perceptually linearly} during lighting changes---faster to light-to-dark convergence reads perceptually as a linear drop in brightness. We determined experimentally that an exponent of 5.0f leads to best results (lower does not converge as fast, higher does not converge any faster). See our video supplement for results. Code listing is give in Figure ~\ref{fig:irradiancegammapseudocode}.
\vspace{-1.5mm}

\begin{figure}
\begin{lstlisting}
float irradianceGamma = 5.0f;

// Perception encoding during probe update

// Passed in or computed earlier in the shader
in vec3 sumOfCosineWeightedRayContributions;
in vec3 oldValue;
in float hysteresis;

vec3 newIrradiance = 
	pow(sumOfCosineWeightedRayContributions, invIrradianceGamma);

return lerp(newIrradiance, oldValue, hysteresis);

//////////////////////////////////////

// Perception decoding during probe sampling
vec3  irradiance = vec3(0);
// For the 8 probes in the surrounding cage
for (int i = 0; i < 8; ++i):
	vec3 probeIrradiance = texture(irradianceTexture, texCoord).rgb;
	// Decode the tone curve, but leave a gamma = 2 curve
	// to approximate sRGB blending for the trilinear
	probeIrradiance = pow(probeIrradiance, 
	vec3(irradianceGamma * 0.5));
	
	irradiance += probeWeight * probeIrradiance;

// Go back to linear irradiance
irradiance = square(irradiance);

return irradiance;
\end{lstlisting}
   \caption{\label{fig:irradiancegammapseudocode}
Perceptual encoding and decoding of probe irradiance during update and sampling.}
\end{figure}

This perception-based encoding has the additional effect of reducing low frequency flicker due to fireflies---bright flashes in the diffuse GI caused by an update ray hitting a small, bright irradiance source.

 \subsection{Fast Convergence Heuristics}
\label{sec:heuristics}
We further accelerate convergence with new heuristic based on per-texel thresholding for irradiance data. Our lower threshold detects changes with magnitude above 25\% of maximum value and lowers the hysteresis by 0.15f. Our higher threshold detects changes with magnitude above 80\% and lowers the hysteresis to 0.0f---we assume in this case that the distribution the probe is sampling has changed completely. These thresholds are active only for irradiance updates---we found them to be too unstable when updating visibility. See Figure ~\ref{fig:probeupdatepseudocode}.

We also implement scene-dependent, per-probe heuristics that adjust the hysteresis based on lighting or geometry changes. These are as follows:

\begin{itemize}
\item Small lighting change (e.g. player-held flashlight turns on): reduce irradiance hysteresis by 15\% for 4 frames.
\item Large lighting change (e.g. abrupt time of day shift): reduce irradiance hysteresis by 50\% for 10 frames.
\item Large object change (e.g. ceiling caves in): reduce irradiance hysteresis by 50\% for 10 frames and visibility hysteresis by 50\% for 7 frames.
\end{itemize}

In all our heuristics, we try to avoid low hysteresis for visibility updates as much as possible to achieve the most stable result. In each of the scene dependent heuristics, hysteresis for \emph{all} probes (not just the probes local to the change) is reduced.

Many effective heuristics exist for adjusting probe hysteresis per-texel and per-probe on a scene dependent basis---we have not explored this space in depth. For example, it would probably be more effective to reduce hysteresis only for probes affected by a lighting or object change rather than for all probes in the scene. While exploring more specific and sensitive heuristics remains a fruitful subject for future work, the heuristics presented here worked well enough for us as we integrated the technique into multiple engines. We never came across content that forced us to adapt them, but our survey was not exhaustive.

\newpage
\begin{figure}
\begin{lstlisting}
// Irradiance Probe Update With Per-Texel Hysteresis Adjustment

// Sum ray contributions
in vec3 sumOfCosineWeightedRayContributions;
in vec3 oldValue;

in float hysteresis;

const float significantChangeThreshold = 0.25;
float newDistributionChangeThreshold = 0.8;


float changeMagnitude = maxComponent(result.rgb - oldValue.xyz);

// Lower the hysteresis when a large change is detected
if (abs(changeMagnitude) > significantChangeThreshold)
    hysteresis = max(0, h - 0.15);

if (abs(changeMagnitude) > newDistributionChangeThreshold)  {
    hysteresis = 0.0f;
}

return lerp(sumOfCosineWeightedRayContributions, oldValue, h);
\end{lstlisting}
   \caption{\label{fig:probeupdatepseudocode}
Pseudocode for probe update with per-texel hysteresis adjustment.}
\end{figure}

Note that temporal anti-aliasing (TAA) applies its own hysteresis, so the base hysteresis for our technique can be lower if TAA is applied. In this case, the TAA hysteresis should be adjusted according to scene heuristics just like the probe hysteresis, or else it will always add a large cost to convergence even on a dramatic lighting or object change.

 \subsection{Second Order Glossy}

\begin{figure}[htb]
  \centering
   \includegraphics[width=0.45\columnwidth]{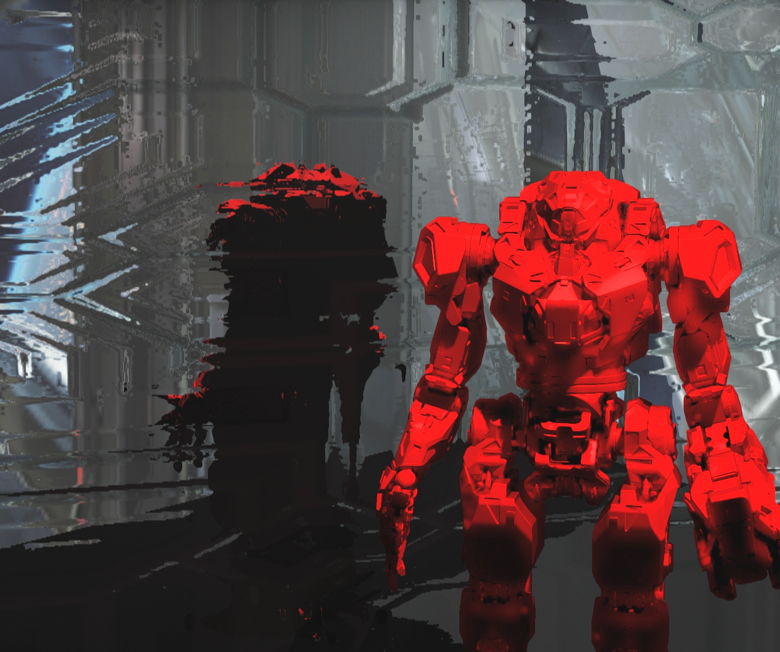}
   \includegraphics[width=0.45\columnwidth]{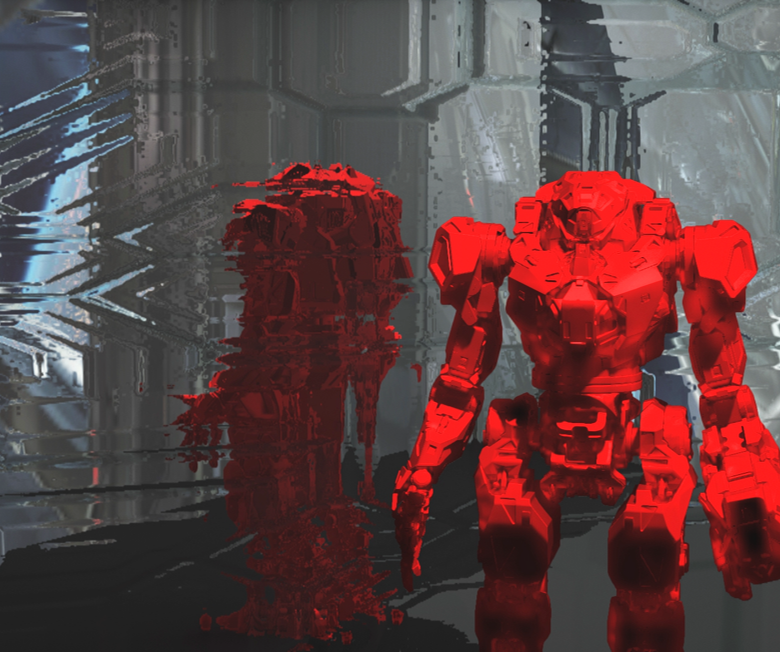}
   \caption{\label{fig:secondorderglossy}
   A shiny robot against a mirror background. Both the mirror background and the robot have high glossy reflectance. The left image shows no second order glossy reflections, while the right image shows second order glossy reflections sampled from probes. }
\end{figure}

We compute glossy reflections with a half-screen resolution wavefront ray trace. These shaded ray hits are then blurred according to surface roughnes and distance from the camera before being integrated into the indirect radiance computation during the deferred shading pass. These raytraced reflections are more realistic than screen-space reflections, but tracing rays for 2nd through nth order reflections is infeasible on most scenes. We improve reflections by reusing the filtered radiance data in the probess to shade 2nd through nth order glossy reflections, resulting in better image quality with minimal performance overhead. See Figure~\ref{fig:secondorderglossy} for an on/off comparison.

 It is common practice in production path tracing to reduce noise by roughening surfaces (or otherwise truncating the BSDF evaluation) on recursive bounces~\cite{PTP2019}. Reusing the irradiance probes for second order reflections is a similar approximation, which here avoids noise by taking advantage of a data structure already available to us. Note, however, that the probe data structure stores cosine-filtered irradiance---not the cosine-weighted \emph{integral} of radiance over the hemisphere, which is the correct measure for reflectance. These two quantities are equivalent to a factor of $2 \pi$, but the units are different: radiance ($\textrm{Ws}^{-1}\textrm{m}^{-2}$) vs. irradiance ($\textrm{Wm}^{-2}$).

\section{Probe Position Adjustment}
\label{sec:OP}
\begin{figure}[htb]
  \centering
   \fbox{\includegraphics[width=0.5\columnwidth]{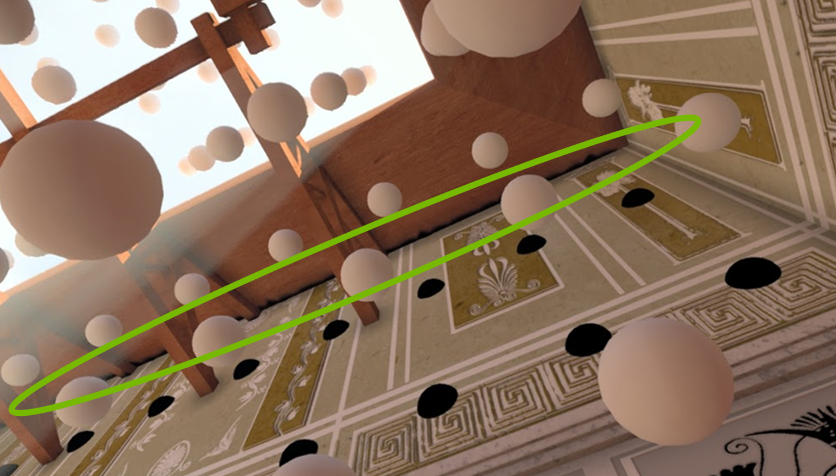}}~\fbox{\includegraphics[width=0.5\columnwidth]{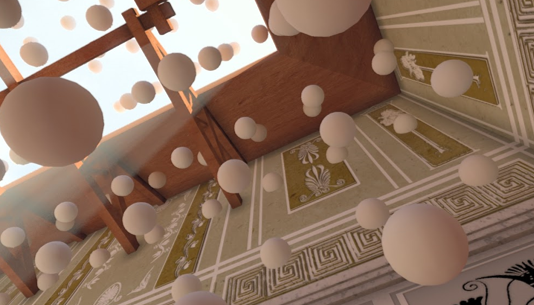}}
   \caption{\label{fig:beforeafteroptimizer}
A view of the ceiling on our Greek Villa scene. Spheres are a visualization of the probes. The black probes are correctly dark, but are not contributing to the final image. The acute corner leads to shadow leaking (labeled with a green ellipse) with a default probe grid (left). Our optimizer adjusts probes out of the wall and ceiling to remove the leak (right).    
}
\end{figure}

The probe visibility information prevents light and shadow leaks from occluded probes, but leaves some probes in total occlusion such that they never contribute to shading. We present a simple, fast optimizer that iteratively shifts probes around static geometry to maximize the number of useful probes and generate good viewpoints. During initialization, our optimizer adjusts each probe through the closest backface it can see, then further adjusts probes \textit{away} from close front-faces to maximize surface visibility (see Figure~\ref{fig:beforeafteroptimizer}). Pseudocode is given in Figure ~\ref{fig:optimizerpseudocode}. We do not move probes around dynamic geometry because this causes instability---a stable result is preferable to an unstable result with lower average error. To correctly light dynamic objects, we leverage the fact that a \emph{uniformly} sampled probe is an approximation of the full irradiance field at its sample location. If a probe passes through a dynamic object, our backface heuristics (described at the end of Section~\ref{sec:selfshadowbias}) will minimize shadow leaking. When the probe emerges, our convergence heuristics (Section~\ref{sec:heuristics}) will quickly converge its value.

Out of a desire to maintain a uniformly sampled irradiance field representation, we did not implement more complex probe sampling techniques, such as importance sampling, which might speed probe convergence at the cost of stability and on-the-fly generalization to moving geometry. Exploring these update techniques in detail is promising for future work.

\begin{figure}
\begin{lstlisting}
in int backfaceCount; // number of rays that hit backfaces
in vec3 closestBackfaceVector; // direction to closest backface
in vec3 farthestFrontfaceVector; // direction to farthest frontface
in vec3 closestFrontfaceVector; // direction to closest frontface

inout vec3 currentOffset; // Current offset from the grid for this probe.

vec3 fullOffset = vec3(inf);
vec3 offsetLimit = ddgiVolume.probeOffsetLimit * 
ddgiVolume.probeSpacing;

// If there's a close backface AND you see more than 25\% backfaces, 
// assume you're inside something.
if ((float(backfaceCount) / RAYS_PER_PROBE) > 0.25f) {
    // Solve for the maximum scaling possible on each axis.
vec3 positiveOffset = (-currentOffset.xyz + offsetLimit) 
				/ closestBackfaceDirection;
vec3 negativeOffset = (-currentOffset.xyz - offsetLimit) 
				/ closestBackfaceDirection;
vec3 combinedOffset = 
vec3(max(positiveOffset.x, negativeOffset.x), 
             max(positiveOffset.y, negativeOffset.y), 
	  max(positiveOffset.z, negativeOffset.z));
     // Slightly bias this point to ensure we stay within bounds.
    const float epsilon = 1e-3; // Millimeter scale
    float scaleFactor = 
(min(min(combinedOffset.x, combinedOffset.y), 
        combinedOffset.z) - epsilon);

    // If we can't move through the backface, don't move at all.
    fullOffset = currentOffset.xyz + closestBackfaceDirection * 
((scaleFactor <= 1.0f) ? 0.0f : scaleFactor);
} else if (!(dot(farthestDirection, randomOrientation * 
sphericalFibonacci(closestFrontfaceIndex, RAYS_PER_PROBE)) > 0.5f)) {
    // The farthest frontface is also the closest if the probe can
    // only see one surface. In this case, don't move the probe.

    // Move minimum distance possible.
    vec3 farthestDirection = min(0.2f, farthestFrontfaceDistance) * 
    normalize(randomOrientation * 
    sphericalFibonacci(farthestFrontfaceIndex, RAYS_PER_PROBE));

    fullOffset = currentOffset.xyz + farthestDirection;
}

currentOffset = fullOffset;

 
\end{lstlisting}
   \caption{\label{fig:optimizerpseudocode}
Pseudocode for an iteration of the probe position optimizer operating on a single probe.}
\end{figure}

\begin{figure}[htb]
  \centering
   \fbox{\includegraphics[width=\columnwidth]{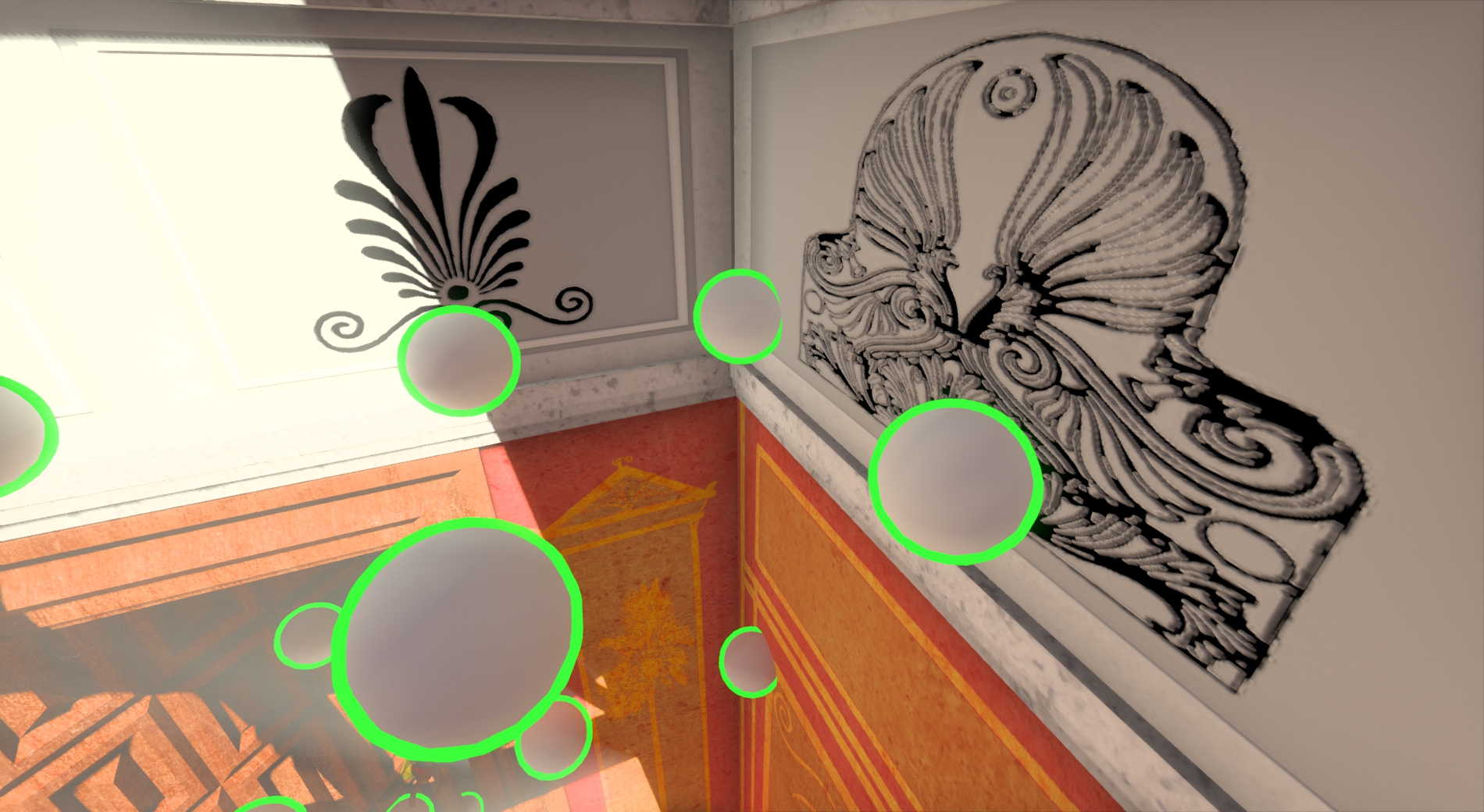}}
   \caption{\label{fig:optimizerdoublecover}
   A corner of the Greek Villa scene. Spheres are visualizations of the probes, encircled in green to denote the "Vigilant" state. Probes are marked "Vigilant" when the optimizer adjusts them out of surfaces, leading to double coverage of surfaces when all 8 probes of a cage can see the front face of the point they're shading.}
\end{figure}

The purpose of the optimizer is to increase the number of probes that can contribute to the final image. The following scenario, however, demonstrates that our optimizer can sometimes add additional computation without increasing image quality. Consider an 8-probe cage surrounding a flat wall (Figure~\ref{fig:optimizerdoublecover}). The optimizer can cause probes to ``double cover" a surface if the 4 probes within the surface are adjusted outside it. This causes the full probe cage to turn on and shade the surface, increasing the number of actively tracing probes without appreciably affecting the image quality (Figure~\ref{fig:optimizerdoublecover}). For our test scenes, this slight inefficiency was worth the added benefit of optimizing probe positions globally.

The probe position optimizer runs for 5 iterations during probe state classification, which is enough for almost all probes to converge their locations. We cap the number of iterations at 5 to prevent probes from moving back and forth (infinitely) through tangent backfaces.

More work is needed to determine the best position optimizer algorithm, and many investigations in this vein exist (see, for example, Wang et al.~\shortcite{FastNonUniformPlacement19}). Our optimizer worked well for multiple engines, but is almost certainly not optimal.

\newpage

\section{Probe States}
\label{sec:probestates}
\begin{figure}[htb]
  \centering
   \includegraphics[width=0.98\columnwidth]{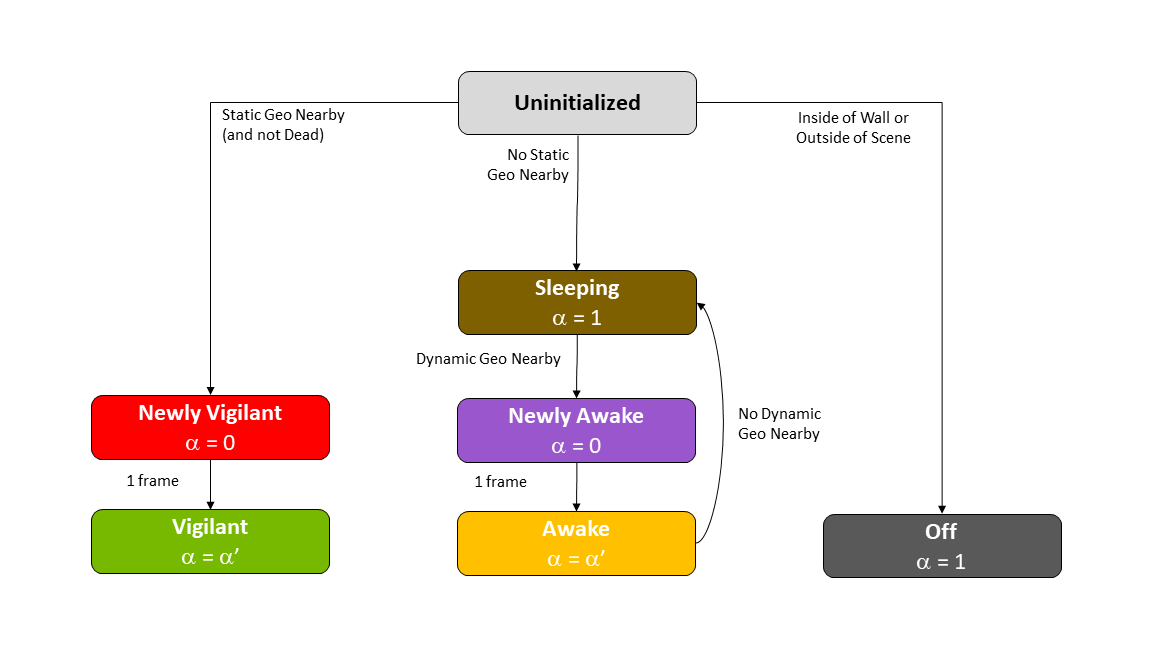}
\vspace{-3mm}
   \caption{\label{fig:probestates}
   Probe states with transitions between each state. $\alpha$ is the hysteresis for the current frame. $\alpha'$ is the default hysteresis for the scene.
}
\end{figure}

For all but the most basic scene geometry,  even after adjustment many probes in a uniform 3D grid will not contribute to the final image. We introduce a robust set of probe states to avoid tracing or updating from such probes to increase performance with the same visual result. Our probe states separate probes that should not update from probes that must, with an additional intermediate state to identify probes that have just appeared (either at scene initialization or with a moving volume---see Section~\ref{sec:cameratrackingvolume}) and adjust their hysteresis accordingly. The full set of states is shown in Figure~\ref{fig:probestates} and discussed in the following sections.

\subsection{Off Probes}
As noted above, the constraints on probe movement imposed by the 3D grid indexing make it impossible to move all probes out of walls (some probes are too constrained by the grid structure). We identify probes that remain inside static geometry and turn them ``Off" (never trace or update). As the optimizer only considers static geometry, probes that happen to spawn inside dynamic geometry are unaffected, and will correctly turn on when appropriate.

\subsection{Probe Update States}
Even probes that are outside static geometry are not used for shading every frame: when no geometry is within $probeSpacing$ of a probe, that probe's value is not used. We set these probes to ``Asleep" and wake them up when a surface is about to use them for shading. Note that a probe needs to be ``Awake" \emph{if and only if} it is shading a surface or about to shade one. Lighting changes and camera proximity do not matter if the probe is not shading a surface. The same is true for making probes ``Asleep": when the camera can't see a probe, it still needs to be ``Awake" if it is shading a surface because it is propagating diffuse irradiance (with 2nd through nth order visibility). Thus, probes that shade static geometry should be ``Vigilant" (they should \emph{always} trace and update). Though probes near geometry must trace to propagate GI, the grid resolution need not be as fine in regions that are far from the camera. Pseudocode for the probe state optimizer is given in Figure ~\ref{fig:probestatepseudocode}.

\paragraph{Participating Media and Probe States} The probe data structure encodes a 3D irradiance field that is queryable at any point within its volume. Thus, it might be queried at positions in empty space to provide global illumination in participating media. In this case, even probes not shading a surface would need to be "Awake" if they are within the participating medium.

\subsection{Full Probe Initialization Algorithm}
\label{sec:fullprobeinit}
Probe positions and states are computed in a four step pass:
\begin{itemize}
\item For all uninitialized probes, trace rays for five frames to determine optimal positioning and initial state. At the end of this pass, all previously uninitialized probes are ``Newly Vigilant", ``Off", or ``Sleeping".
\item Extend AABBs for all dynamic objects by a probe grid cell + the self-shadow bias for a conservative estimate. Set all ``Sleeping" probes inside the extended AABB of a dynamic object to ``Newly Awake".
\item Optionally trace a large number of rays for ``Newly Vigilant" and ``Newly Awake" probes to converge them in a frame, setting hysteresis to 0. Set their states to ``Vigilant" and ``Awake" respectively.
\item Trace rays from ``Vigilant" and ``Awake" probes to update their values with the normal hysteresis value for the scene. This step can also be used to converge ``Newly Vigilant" and ``Newly Awake" probe values if the previous step was omitted.
\end{itemize}

The first step of the algorithm can be greatly accelerated with static geometry bounding boxes, as a probe can be directly adjusted against those bounding boxes rather than relying on distance and backface information from the spherical ray cast. Many probes could be immediately classified ``Newly Vigilant" with this approach, though ray tracing would still be necessary to correctly determine which probes should be set to ``Off".

\begin{figure}
\begin{lstlisting}
for each uninitialized probe:
	Trace rays (distance only, no shading)
	Position optimizer iteration
	if (still in wall):
		OFF
	if (frontfaceDistance < probeSpacing):
		NEWLY VIGILANT
	else
		SLEEPING

for all dynamic geo:
	Extend bounding boxes grid cell size + self shadow bias
	for all SLEEPING probes:
		if (probe inside bounding box):
			NEWLY AWAKE

// Optionally converge probes in this frame...
for all NEWLY AWAKE and NEWLY VIGILANT probes:
	Trace rays to converge value
	NEWLY AWAKE -> AWAKE
	NEWLY VIGILANT -> VIGILANT

// ...or let them converge in the update pass.
for all VIGILANT or AWAKE probes:
	Trace rays and update value.
\end{lstlisting}
   \caption{\label{fig:probestatepseudocode}
Pseudocode for probe state computation. }
\end{figure}

Though these passes run every frame, for the majority of frames the first step will not run because no probes will be uninitialized. If the optional convergence pass is omitted, then only the final update step will run for most frames.

\subsection{Probe Sleeping Performance}
\begin{figure}[htb]
\vspace{-8mm}
  \centering
\begin{subfigure}{0.9\columnwidth}
   \includegraphics[width=\columnwidth]{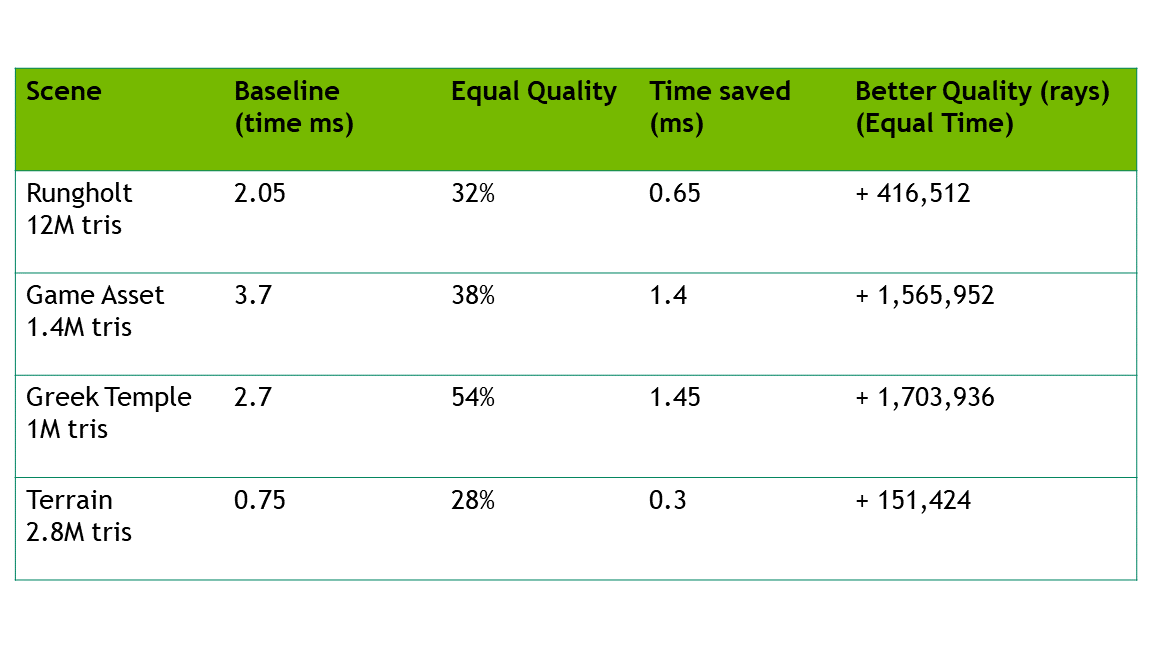}
\end{subfigure}
\vspace{-3mm}
   \caption{\label{fig:probestateperf}
   Performance data for probe sleeping. The "Baseline" column shows the time for probe trace and update without probe sleeping (all probes are marked "Vigilant"). The "Equal Quality" and "Time Saved" columns show savings of probe sleeping as a percentage of time and as absolute time respectively. Finally, the "Better Quality" column shows the absolute ray increase achievable by tracing more rays from active probes to match the baseline time. Higher quality is achieved here by tracing more rays per probe per update pass---this reduces the variance in the estimation and speeds convergence.  }
\end{figure}

Probe sleeping using our probe state scheme leads to a 30-50\% average performance improvement (Figure~\ref{fig:probestateperf}). In addition to the performance improvement (shown in the middle column) we also show corresponding increases in rays cast per probe for the same performance. Casting more rays per probe makes new probe values more stable and allows for a lower global hysteresis, which makes the GI converge faster.

\section{Quantitative performance improvements}

 \subsection{Probe Update Shader Optimization}

The approach of Majercik et al.~\shortcite{Majercik2019Irradiance} updated probe texels using a pixel shader with a stencil buffer (to avoid processing border texels in the update pass). Border texels were updated in a separate pixel shader pass for correct bilinear interpolation. This approach leverages the graphics hardware for alpha blending results. Despite this, however, faster update can be achieved by using a general purpose GPU (GPGPU) compute operation optimized with GPU compute best practices. We give background and details of this approach below.

Modern GPU architectures dispatch thread groups to cover user-specified compute grid dimensions. All threads in a group execute the same code in parallel, so ensuring that threads do not take different control paths in the code (coherent execution) is vital for performance. By ensuring coherent execution, we achieve a 3x performance improvement in the update pass over the pixel shader approach with careful indexing over thread blocks consisting of an integer number of groups. All group execution is fullly coherent. In addition, we store incoming shaded sample ray hits in shared memory buffers so that all threads can read it in parallel when computing a new probe texel value. 

Previous work showed the effect of probe resolution on image quality and performance. We maintain image quality while selecting probe resolution (8x8 irradiance, 16x16 visibility) for a combination of bandwidth, memory footprint, fast convolution, efficient index computation, and most important: mapping to SIMD instructions (thread lanes on a GPU) for peak occupancy on our target hardware. At powers of two, a probe can be updated by an integer number of 32 or 64 thread groups (common hardware-defined minimum sizes) for maximum possible occupation and coherence. Arbitrary resolution values offer the highest flexibility at the cost of efficiency.

\clearpage
\begin{figure}[h!]
\vspace{-5mm}
\begin{subfigure}{\columnwidth}
\centering
\includegraphics[width=0.4\columnwidth]{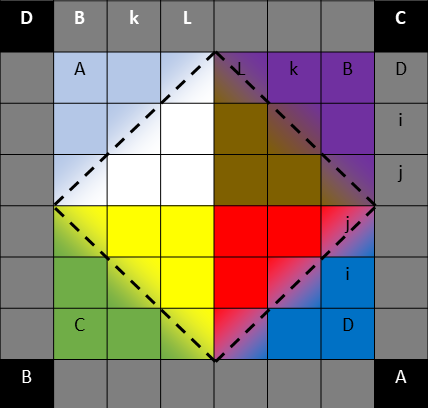}
\caption{Octahedral representation and border copy texels. Colors denote faces on the collapsed octahedron. Letters in border cells denote copy destinations for cells inside the border labeled with the same letter.}
\end{subfigure}
\begin{subfigure}{\columnwidth}
\centering
\includegraphics[width=0.7\columnwidth]{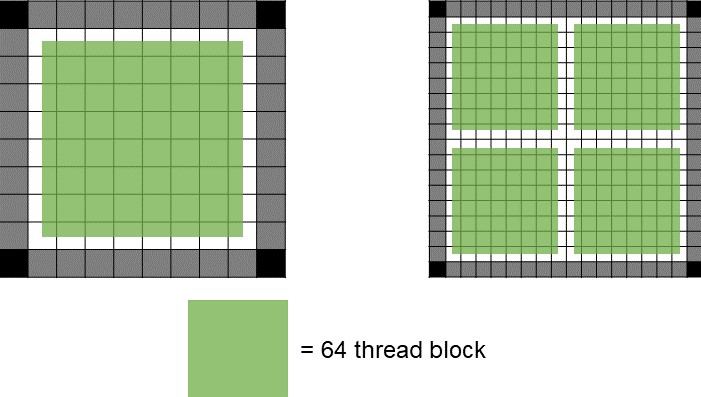}
\caption{Thread block alignment for probe update on an 8x8 irradiance probe (left) and a 16x16 visibility probe (right).}
\end{subfigure}
\begin{subfigure}{\columnwidth}
\includegraphics[width=\columnwidth]{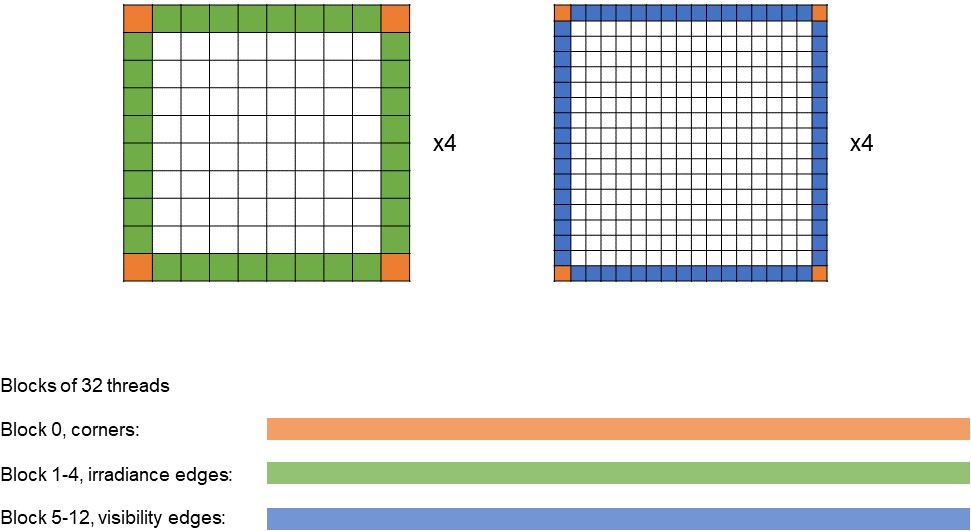}
\caption{Thread block alignment for probe border copy. One block of 32 threads copies corners for four irradiance and four visibility probes (orange). Four blocks copy edges for four irradiance probes (green). Eight blocks copy edges for four visibility probes (blue).}
\end{subfigure}
   \caption{\label{fig:probediagram}
    Octahedral probe layout and probe update thread indexing.}
\end{figure}
\clearpage

Figure~\ref{fig:probediagram} shows details of our compute shader indexing, including an example octahedral probe encoding to illustrate border-texel copy for correct hardware bilinear interpolation. Our optimized compute shader is included alongside the update shader of the previous technique~\shortcite{Majercik2019Irradiance} in the supplemental material.

\subsection{Tracking Windows}
\label{sec:cameratrackingvolume}
\begin{figure}[htb]
  \centering
\begin{subfigure}{0.49\columnwidth}
   \hspace*{-.85cm}\includegraphics[width=\columnwidth]{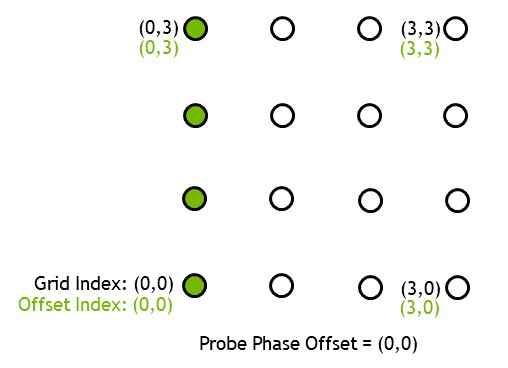}
\caption{Default Grid}
\end{subfigure}
\begin{subfigure}{0.34\columnwidth}
   \includegraphics[width=\columnwidth]{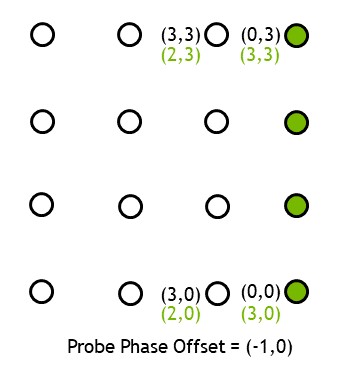}
\caption{Offset Grid}
\end{subfigure}
   \caption{\label{fig:cameraIndexing}
    Conceptual layout of the camera tracking window indexing with phase offset in 2D. The row of probes that moves is colored in green. When the camera passes the center bounding threshold moving in the +X direction, the leftmost row of probes leapfrogs to the +X face of the volume. The newly computed grid index is shown in green. The corresponding phase offset change is shown on the right. }
\end{figure}

Conceptually, a probe grid covers all space in the scene. In practice, however, we do not have the compute or raytracing budget to update and trace a level-sized, high resolution probe grid as it may contain tens of thousands of probes. To maintain high probe resolution where it is most necessary, we implement a 3D tracking window of probes. We used this window to track the camera, though any object can be tracked with the same strategy. Our window begins centered on the camera. As the camera moves, if it moves further from the center than the distance between two probes in a cage (along any axis),  a new plane of probes spawns in front of it (relative to its direction of motion) and the plane furthest behind it disappears. We implement this behavior using a 3D fixed-length circular buffer. When a new probe plane appears and is initialized, its new values are written to the memory of the plane in the last row behind the camera: the probes "leapfrog" over the camera in discrete steps (Figure~\ref{fig:cameraIndexing}). A discretely stepping probe window necessitates careful interpolation between multiple probe volumes---our strategy for this is discussed in Section~\ref{sec:multipleprobevolumes}.

\subsection{Multiple Probe Volumes}
\label{sec:multipleprobevolumes}
Multiple probe volumes at differing resolutions can be used to efficiently implement progressively decreasing grid resolutions that cascade out from the camera, thus saving performance without effecting image quality. The data for these probe volumes is packed into a single texture as shown in Figure ~\ref{fgi:multivolumepacking} The same approach is used in geoclipmaps \cite{GeoClipmap}, light propagation volumes \cite{LightPropagationVolumes2010}, and voxel cone volumes \cite{Crassin2011Voxel}. Additional high-resolution volumes can also be used to efficiently cover hero assets with complex geometry that require higher resolution diffuse irradiance.

\begin{figure}[h]
  \centering
\begin{adjustwidth}{-2.5cm}{}
\begin{subfigure}{0.45\columnwidth}
   \includegraphics[width=\columnwidth]{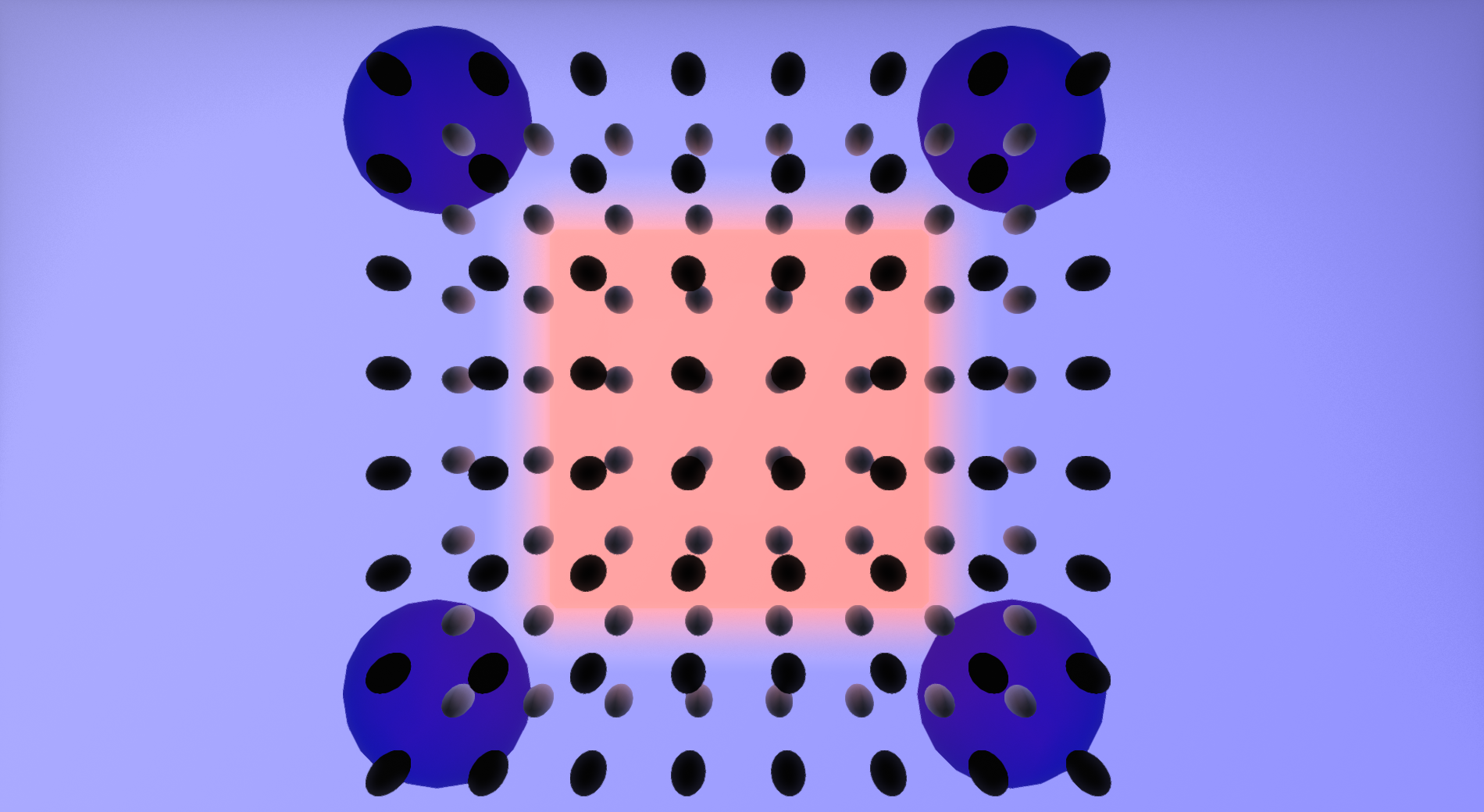}
\caption{Multiple probe volumes}
\end{subfigure}~\begin{subfigure}{0.45\columnwidth}
   \includegraphics[width=\columnwidth]{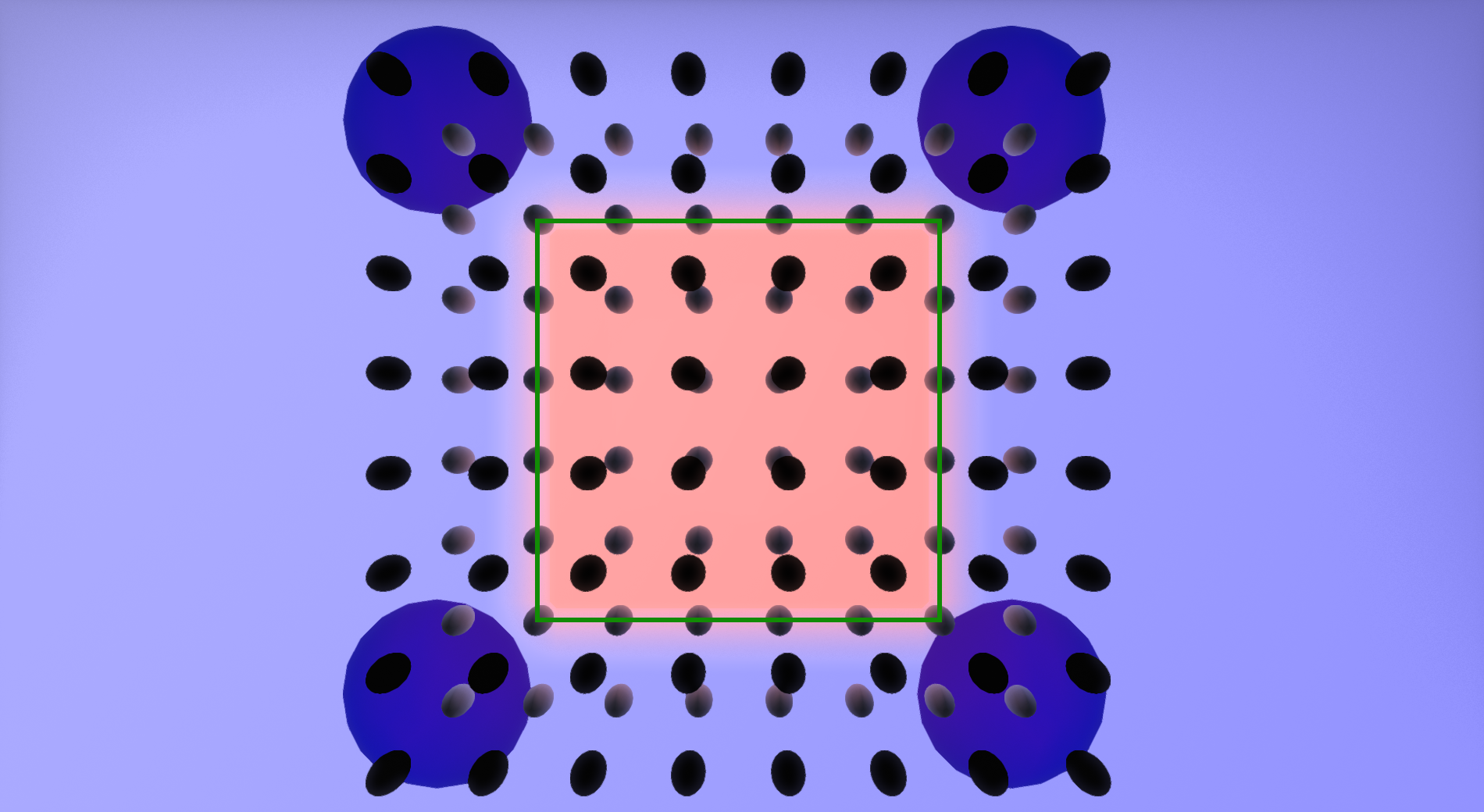}
\caption{Transition start (marked in green)}
\end{subfigure}~\begin{subfigure}{0.45\columnwidth}
   \includegraphics[width=\columnwidth]{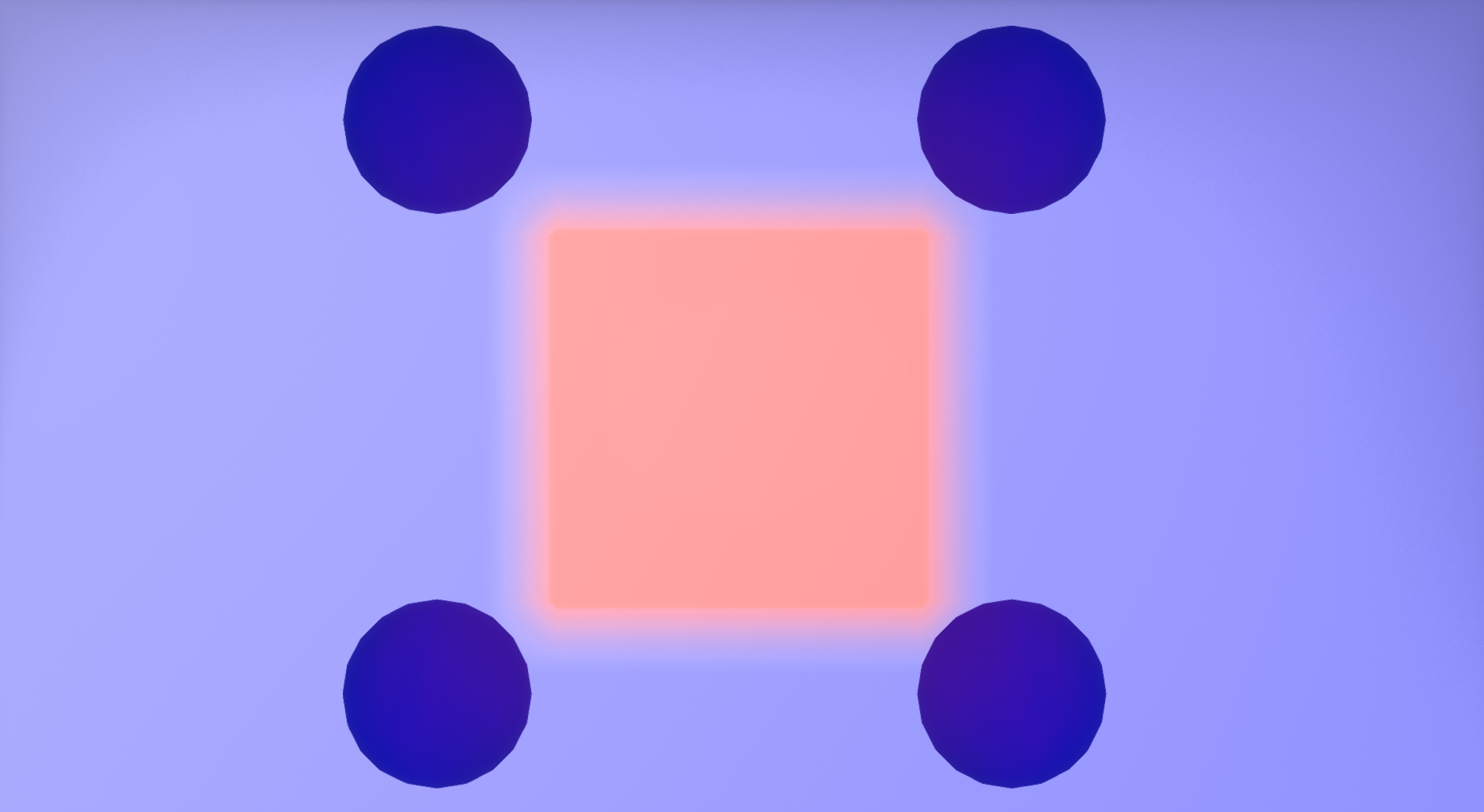}
\caption{Dense volume hidden}
\end{subfigure}
\end{adjustwidth}
   \caption{\label{fig:multivolumeblending}
   Spheres visualized show a dense volume (smaller sphers) and a sparse volume (larger spheres). The spheres are sized based on the probe spacing within each volume. On the far left, the pink region shows the the area fully shaded by the dense volume, which gradually falls off to blue, the area shaded by the sparse volume. The center image marks the start of this transition. The rightmost image hides the dense probes to make visualizing the transition region easier.}
\end{figure}

\begin{figure}[htb]
\centering
\begin{adjustwidth}{-2.55cm}{}
\includegraphics[width=1.37\columnwidth]{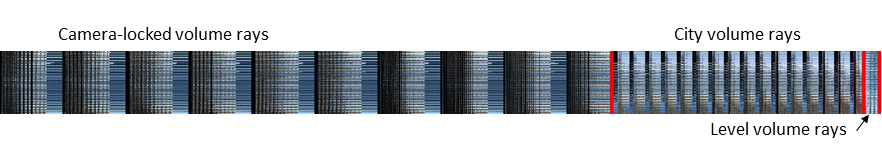}
\end{adjustwidth}
   \caption{\label{fig:multivolumepacking}
    Shaded ray hit data for multiple volumes packed into a single texture. This texture is irradiance data taken from our multivolume scene in the supplemental video. The texture includes shaded update rays for the camera locked volume, the city scale volume, and the level scale volume---these are labeled in the figure and delineated within the texture by the red lines (which are not part of the irradiance data).}
\end{figure}

We blend between volumes by linearly falling off from 1.0-0.0 at the last grid cell (starting at the second-to-last plane of probes) along each axis of the 3D grid (see Figure~\ref{fig:multivolumeblending}). In the deferred shader, a weight is computed for each volume starting from most to least dense. This is also the sampling order because the most dense volume will have the best approximation of the local lightfield. Volume weights are accumulated at each volume sample. After the weight total reaches 1.0, further volumes are skipped.

\begin{figure}[h!]
\vspace{-5mm}
  \centering
\begin{subfigure}{\columnwidth}
   \includegraphics[width=0.75\columnwidth]{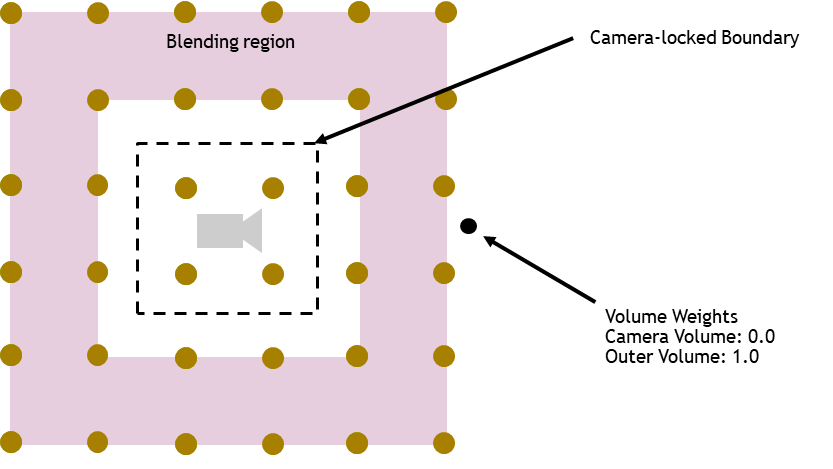}
\begin{adjustwidth}{0cm}{}
\caption{Initial camera position. Labels show the blending region for the camera tracking window, the camera boundary that will cause the volume to move, and the volume weights for a point being shaded by the camera volume (brown circles) and a surrounding volume (not visualized).}
\end{adjustwidth}
\end{subfigure}
\begin{subfigure}{\columnwidth}
   \includegraphics[width= 0.75\columnwidth]{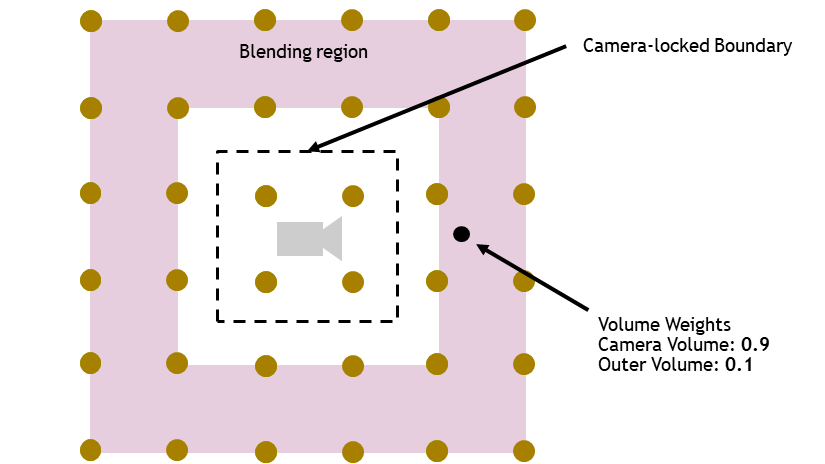}
\begin{adjustwidth}{0cm}{}
\caption{Camera moves. Without camera-aware blending, volume weights on the point change dramatically in one frame.}
\end{adjustwidth}
\end{subfigure}
\begin{subfigure}{\columnwidth}
   \includegraphics[width=0.75\columnwidth]{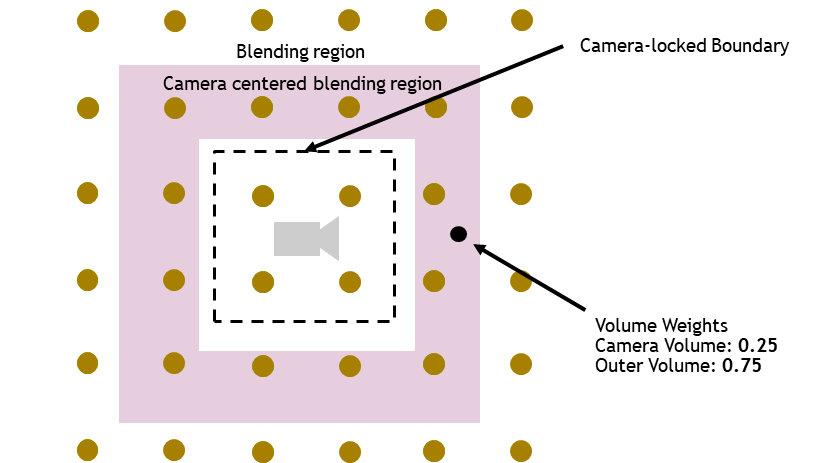}
\begin{adjustwidth}{0cm}{}
\caption{Camera moves. With camera-aware blending, the volume weights change slowly over the course of multiple frames, leading to smoother transtions.}
\end{adjustwidth}
\end{subfigure}
\vspace{-5mm}
   \caption{\label{fig:cameraVolumeBlending}
   2D illustration of volume blending using the static volume method vs. a camera aware volume blending. }
\end{figure}

The weighted volume blending described above yields smooth transitions for static volumes, but can cause popping in the GI when applied to camera locked volumes. When a volume leapfrogs in front of the camera, some points can go from being fully shaded by a sparse cascade to being heavily shaded by the camera cascade (Figure~\ref{fig:cameraVolumeBlending}). When computing blending weights for camera locked volumes, we address this by tightening the transition region by one grid cell (along each axis) then centering it on the camera. When a new plane of probes leapfrogs to the front of a volume, points that are newly within that volume will not immediately be shaded by it. Instead, those points will gradually transition between volumes as the camera moves towards them. Results are shown in our supplemental video.

The prototype multivolume code passes all probe volumes to the deferred shader, and then per-pixel iterates through them to figure out which ones contain the point being shaded. Though not the optimal approach for performance, this provides the highest flexibility in tweaking the blending algorithm to evaluate image quality. For a production implementation, the usual solutions for the deferred shading light loop issue (considering the volumes as lights) are available:
\begin{itemize}
\item Do the full brute force light loop---for fewer than 10 volumes, the point-in-OBB test to determine which volumes contain the shaded point is fast to evaluate.
\item Make one deferred pass per volume, rasterizing the volume's bounds to find the covered pixels.
\item Make a spatial data structure (e.g., octtree, BVH) over the volumes and then traverse that at runtime in the pixel shader to find which volumes the pixel is in. This method requires more bookkeeping and potentially costly data-dependent fetches.
\item Use tiles~\cite{TiledShading12} set up on the CPU or with a GPU pass to conservatively approximate one of the previous methods.
\end{itemize}

For the pure cascaded method, these optimizations are not necessary because volumes are axis-aligned in world space and nested in a regular pattern.

 \subsection{Inline Shading}
Previous probe schemes required an extra shader pass to gather the indirect contribution over the frame. We present a simpler framework that optimizes the global illumination gather step to directly sample the probe data structure during shading, yielding reduced bandwidth requirements. Our code is included in the supplemental material in \lstinline{GIRenderer_deferredShade.pix}.

\clearpage
\clearpage

\section{Conclusion and Discussion}
\label{sec:conclusion}

We present multiple extensions to the dynamic diffuse global illumination algorithm~\cite{Majercik2019Irradiance} to improve image quality, performance, and ease of deployment in a production setting. These extensions were developed in response to production constraints encountered when integrating the technique into the NVIDIA RTXGI SDK~\cite{RTXGISDK}, the Unity game engine, Unreal Engine 4, and several commercial games.

The base algorithm of Majercik et al.~\shortcite{Majercik2019Irradiance} is inherently practical due to it's image quality and performance. This paper covers the gap between a practical algorithm and one that is ready for production deployment. Extensions like our ``self-shadow bias" make the algorithm easier to tune, and our performance optimizations to the update pass make it feasible for the render budget of production games. For all of our extensions, we sought solutions that were robust, easy to understand, and easy to tune without fundamentally changing the algorithm.

\subsection{Limitations and Future Work}

Though our proposed convergence heuristics increase convergence over the previous approach, there is still some ghosting in the indirect illumination for small, bright light sources (like flashlights---see our video supplement at 7:05). This lag could be addressed by intensifying our specific hysteresis-reduction heuristics on small lights known to cause ghosting, though doing this globally may cause instability in other regions of the image. While more specialized methods like reflective shadow maps yield less ghosting~\cite{ReflectiveShadowMaps2016}, an advantage of our method is that all light sources can be handled generically to produce global illumination---we trade some quality for generality.

In addition to our performance improvements, a per-frame ray budget could be implemented to allow more control over the render budget of the technique. For our applications, we found that controlling a) the rays per probe and b) the number of probes in a volume was enough to hit our performance targets. A more sophisticated treatment of ray budget would trace different ray amounts on a per-probe basis, adding a lot of complexity to the implementation. We chose simplicity over a more optimized ray budget, but a study of optimal ray apportioning between probes (taking into account lighting and geometry changes, the camera position, etc.)  is interesting future work.

Our algorithm covers a large space of rendered effects and thus suggests many possible directions for future work. For instance, our techniqe forces second-order glossy reflections to maximum roughness in order to re-use the irradiance values as cosine-filtered radiance. Increasing roughness over scattering events has precedent as a noise reduction technique in film production~\cite{SonyArnold2018}, although typically not at such an aggressive scale.

Second order glossy reflections could be improved by using multiple higher resolution filtered radiance textures with different cosine power weighting---like the weighting for visibility probes, but with multiple octahedral representations per sample point instead of one. These could be used to render second order glossy reflections of varying roughness.

\subsection*{Acknowledgements}
 Foremost, we thank Peter Shirley for his invaluable feedback and editing.
Thanks to Corey Taylor and Mike Mara for the initial probe implementation. Thanks to Derek Nowrouzezahrai and Jean-Philippe Guertin for their effort in the original DDGI paper. Thanks to Paul Hodgson, Peter Featherstone, Jesper Mortensen, Kuba Cupisz, and the rest of the Unity Copenhagen lighting team for their help with Unity. Thanks to Kelsey Blanton and Alan Wolfe for their work on the NVIDIA RTXGI SDK. Thanks to Pablo Palmier at Ninja Theory for his help with Unreal Engine 4. 

\small
\bibliographystyle{jcgt}
\bibliography{paper}

\begin{thebibliography}{\protect\citename{Kaplanyan and Dachsbacher }2010}

\bibitem[\protect\citename{Cigolle et~al\mbox{.} }2014]{Cigolle2014Vector}
{\sc Cigolle, Z.~H., Donow, S., Evangelakos, D., Mara, M., McGuire, M., and
  Meyer, Q.}
\newblock 2014.
\newblock A survey of efficient representations for independent unit vectors.
\newblock {\em Journal of Computer Graphics Techniques (JCGT) 3}, 2 (April),
  1--30.
\newblock URL: \url{http://jcgt.org/published/0003/02/01/}.

\bibitem[\protect\citename{Crassin et~al\mbox{.} }2011]{Crassin2011Voxel}
{\sc Crassin, C., Neyret, F., Sainz, M., Green, S., and Eisemann, E.}
\newblock 2011.
\newblock Interactive indirect illumination using voxel cone tracing.
\newblock {\em Computer Graphics Forum 30}, 7, 1921--1930.
\newblock URL:
  \url{https://onlinelibrary.wiley.com/doi/abs/10.1111/j.1467-8659.2011.02063.x},
  arXiv:https://onlinelibrary.wiley.com/doi/pdf/10.1111/j.1467-8659.2011.02063.x,
  doi:10.1111/j.1467-8659.2011.02063.x.

\bibitem[\protect\citename{Fascione et~al\mbox{.} }2019]{PTP2019}
{\sc Fascione, L., Hanika, J., Heckenberg, D., Kulla, C., Droske, M., and
  Schwarzhaupt, J.}
\newblock 2019.
\newblock Path tracing in production: Part 1: Modern path tracing.
\newblock In {\em ACM SIGGRAPH 2019 Courses}, Association for Computing
  Machinery, New York, NY, USA, SIGGRAPH ’19.
\newblock URL: \url{https://doi.org/10.1145/3305366.3328079},
  doi:10.1145/3305366.3328079.

\bibitem[\protect\citename{Hooker }2016]{Hooker2016GI}
{\sc Hooker, J.}
\newblock 2016.
\newblock Volumetric global illumination at treyarch.
\newblock In {\em Advances in Real-Time Rendering 2016}, SIGGRAPH 2016,
  Treyarch.
\newblock URL:
  \url{https://www.activision.com/cdn/research/Volumetric_Global_Illumination_at_Treyarch.pdf}.

\bibitem[\protect\citename{Jendersie et~al\mbox{.}
  }2016]{Jendersie_jcgt16_RTGI}
{\sc Jendersie, J., Kuri, D., and Grosch, T.}
\newblock 2016.
\newblock {Real-Time} {Global} {Illumination} {Using} {Precomputed}
  {Illuminance} {Composition} with {Chrominance} {Compression}.
\newblock {\em Journal of Computer Graphics Techniques (JCGT) 5}, 4 (December),
  8--35.
\newblock URL: \url{http://jcgt.org/published/0005/04/02/}.

\bibitem[\protect\citename{Kaplanyan and Dachsbacher
  }2010]{LightPropagationVolumes2010}
{\sc Kaplanyan, A., and Dachsbacher, C.}
\newblock 2010.
\newblock Cascaded light propagation volumes for real-time indirect
  illumination.
\newblock In {\em Proceedings of the 2010 ACM SIGGRAPH Symposium on Interactive
  3D Graphics and Games}, Association for Computing Machinery, New York, NY,
  USA, I3D ’10, 99–107.
\newblock URL: \url{https://doi.org/10.1145/1730804.1730821},
  doi:10.1145/1730804.1730821.

\bibitem[\protect\citename{Keller et~al\mbox{.} }2015]{keller2015path}
{\sc Keller, A., Fascione, L., Fajardo, M., Georgiev, I., Christensen, P.,
  Hanika, J., Eisenacher, C., and Nichols, G.}
\newblock 2015.
\newblock The path tracing revolution in the movie industry.
\newblock In {\em ACM SIGGRAPH 2015 Courses}. 1--7.

\bibitem[\protect\citename{Kulla et~al\mbox{.} }2018]{SonyArnold2018}
{\sc Kulla, C., Conty, A., Stein, C., and Gritz, L.}
\newblock 2018.
\newblock Sony pictures imageworks arnold.
\newblock {\em ACM Trans. Graph. 37}, 3 (Aug.).
\newblock URL: \url{https://doi.org/10.1145/3180495}, doi:10.1145/3180495.

\bibitem[\protect\citename{Lagarde and Zanuttini }2012]{Sebastien2012Parallax}
{\sc Lagarde, S., and Zanuttini, A.}
\newblock 2012.
\newblock Local image-based lighting with parallax-corrected cubemap.
\newblock SIGGRAPH 2012, DONTNOD Entertainment.
\newblock URL:
  \url{https://seblagarde.wordpress.com/2012/11/28/siggraph-2012-talk/}.

\bibitem[\protect\citename{Losasso and Hoppe }2004]{GeoClipmap}
{\sc Losasso, F., and Hoppe, H.}
\newblock 2004.
\newblock Geometry clipmaps: Terrain rendering using nested regular grids.
\newblock {\em ACM Trans. Graph. 23}, 3 (Aug.), 769–776.
\newblock URL: \url{https://doi.org/10.1145/1015706.1015799},
  doi:10.1145/1015706.1015799.

\bibitem[\protect\citename{Majercik et~al\mbox{.}
  }2019]{Majercik2019Irradiance}
{\sc Majercik, Z., Guertin, J.-P., Nowrouzezahrai, D., and McGuire, M.}
\newblock 2019.
\newblock Dynamic diffuse global illumination with ray-traced irradiance
  fields.
\newblock {\em Journal of Computer Graphics Techniques (JCGT) 8}, 2 (June),
  1--30.
\newblock URL: \url{http://jcgt.org/published/0008/02/01/}.

\bibitem[\protect\citename{Martin and Einarsson }2010]{Martin2010Radiosity}
{\sc Martin, S., and Einarsson, P.}
\newblock 2010.
\newblock A real time radiosity architecture for video games.
\newblock In {\em Advances in Real-Time Rendering 2010}, SIGGRAPH 2010,
  Geomerics and EA DICE.
\newblock URL:
  \url{http://advances.realtimerendering.com/s2010/Martin-Einarsson-RadiosityArchitecture(SIGGRAPH
  2010 Advanced RealTime Rendering Course).pdf}.

\bibitem[\protect\citename{McAuley }2012]{FarCry2012}
{\sc McAuley, S.}
\newblock 2012.
\newblock Calibrating lighting and materials in far cry 3.
\newblock In {\em Practical Physically Based Shading in Film and Game
  Production}, SIGGRAPH 2012, Ubisoft Montreal.
\newblock URL:
  \url{https://blog.selfshadow.com/publications/s2012-shading-course/}.

\bibitem[\protect\citename{McGuire et~al\mbox{.} }2017]{Mara17Lightfield}
{\sc McGuire, M., Mara, M., Nowrouzezahrai, D., and Luebke, D.}
\newblock 2017.
\newblock Real-time global illumination using precomputed light field probes.
\newblock In {\em {ACM} {SIGGRAPH} Symposium on Interactive {3D} Graphics and
  Games}, 11.
\newblock URL:
  \url{http://casual-effects.com/research/McGuire2017LightField/index.html}.

\bibitem[\protect\citename{NVIDIA }2020]{RTXGISDK}
{\sc NVIDIA}, 2020.
\newblock Rtx global illumination, Sep.
\newblock URL: \url{https://developer.nvidia.com/rtxgi}.

\bibitem[\protect\citename{Olsson et~al\mbox{.} }2012]{TiledShading12}
{\sc Olsson, O., Billeter, M., and Assarsson, U.}
\newblock 2012.
\newblock Tiled and clustered forward shading: Supporting transparency and
  msaa.
\newblock In {\em ACM SIGGRAPH 2012 Talks}, Association for Computing
  Machinery, New York, NY, USA, SIGGRAPH ’12.
\newblock URL: \url{https://doi.org/10.1145/2343045.2343095},
  doi:10.1145/2343045.2343095.

\bibitem[\protect\citename{Ritschel et~al\mbox{.}
  }2009]{Ritschel:2009:ADG:1507149.1507161}
{\sc Ritschel, T., Grosch, T., and Seidel, H.-P.}
\newblock 2009.
\newblock Approximating dynamic global illumination in image space.
\newblock In {\em Proceedings of the 2009 Symposium on Interactive 3D Graphics
  and Games}, ACM, New York, NY, USA, I3D '09, 75--82.
\newblock URL: \url{http://doi.acm.org/10.1145/1507149.1507161},
  doi:10.1145/1507149.1507161.

\bibitem[\protect\citename{Silvennoinen and Lehtinen }2017]{Silvennoinen:2017}
{\sc Silvennoinen, A., and Lehtinen, J.}
\newblock 2017.
\newblock Real-time global illumination by precomputed local reconstruction
  from sparse radiance probes.
\newblock {\em ACM Transactions on Graphics (Proceedings of SIGGRAPH Asia) 36},
  6 (11), 230:1--230:13.
\newblock URL: \url{https://doi.org/10.1145/3130800.3130852},
  doi:10.1145/3130800.3130852.

\bibitem[\protect\citename{Stachowiak and Uludag
  }2015]{Stachowiak2015Stochastic}
{\sc Stachowiak, T., and Uludag, Y.}
\newblock 2015.
\newblock Stochastic screen-space reflections.
\newblock In {\em Advances in Real-Time Rendering 2015}, SIGGRAPH 2015, EA
  DICE.
\newblock URL:
  \url{https://www.ea.com/frostbite/news/stochastic-screen-space-reflections}.

\bibitem[\protect\citename{Stefanov }2016]{GIDivision}
{\sc Stefanov, N.}
\newblock 2016.
\newblock Global illumination in tom clancy's the division.
\newblock Presented at Game Developers Conference, 2016.
\newblock URL:
  \url{https://www.youtube.com/watch?v=04YUZ3bWAyg&feature=youtu.be&t=657}.

\bibitem[\protect\citename{Toth et~al\mbox{.} }2015]{Toth2015Disparity}
{\sc Toth, R., Hasselgren, J., and Akenine-M\"{o}ller, T.}
\newblock 2015.
\newblock Perception of highlight disparity at a distance in consumer
  head-mounted displays.
\newblock In {\em Proceedings of the 7th Conference on High-Performance
  Graphics}, ACM, New York, NY, USA, HPG '15, 61--66.
\newblock URL: \url{http://doi.acm.org/10.1145/2790060.2790062},
  doi:10.1145/2790060.2790062.

\bibitem[\protect\citename{Valient }2013]{Valient2014SSRR}
{\sc Valient, M.}
\newblock 2013.
\newblock Killzone shadow fall demo postmortem.
\newblock Sony Devstation 2013, Guerilla Games.
\newblock URL:
  \url{https://www.guerrilla-games.com/read/killzone-shadow-fall-demo-postmortem}.

\bibitem[\protect\citename{Wang et~al\mbox{.} }2019]{FastNonUniformPlacement19}
{\sc Wang, Y., Khiat, S., Kry, P.~G., and Nowrouzezahrai, D.}
\newblock 2019.
\newblock Fast non-uniform radiance probe placement and tracing.
\newblock In {\em Proceedings of the ACM SIGGRAPH Symposium on Interactive 3D
  Graphics and Games}, Association for Computing Machinery, New York, NY, USA,
  I3D ’19.
\newblock URL: \url{https://doi.org/10.1145/3306131.3317024},
  doi:10.1145/3306131.3317024.

\bibitem[\protect\citename{Wyman }2005]{Wyman2005Refraction}
{\sc Wyman, C.}
\newblock 2005.
\newblock An approximate image-space approach for interactive refraction.
\newblock {\em ACM Trans. Graph. 24}, 3 (July), 1050--1053.
\newblock URL: \url{http://doi.acm.org/10.1145/1073204.1073310},
  doi:10.1145/1073204.1073310.

\bibitem[\protect\citename{Xu }2016]{ReflectiveShadowMaps2016}
{\sc Xu, K.}, 2016.
\newblock Temporal antialiasing in uncharted 4.
\newblock URL: \url{http://advances.realtimerendering.com/s2016/index.html}.

\end{thebibliography}

\section*{Index of Supplemental Materials}
The supplemental material contains video results for each of our extensions. The video is available here: https://youtu.be/vbJ2aNI94Ho. The supplemental materials contain relevent C++ and shader code for our extensions. Where appropriate, we have included code from Majerick et al.~\shortcite{Majercik2019Irradiance} for comparison.

\end{document}